\documentclass[trackchanges,twocolumn]{aastex7}
\hypersetup{linkcolor=magenta,citecolor=blue,filecolor=cyan,urlcolor=purple}

\begin{document}

\title{Understanding the Complex Morphology of a Coronal Mass Ejection. \\ 
III: Modeling the CME–SIR Interaction Behind a Severe Geomagnetic Storm}

\author[orcid=0009-0009-9799-979X]{Cecilia Mac Cormack}
\affiliation{Heliophysics Science Division, NASA Goddard Space Flight Center, Greenbelt, MD 20770, USA}
\affiliation{The Catholic University of America, Washington, DC 20064, USA}
\email[show]{cecilia.maccormack@nasa.gov}  

\author[orcid=0000-0001-5400-2800]{Abril Sahade} 
\affiliation{Heliophysics Science Division, NASA Goddard Space Flight Center, Greenbelt, MD 20770, USA}
\email[show]{abril.sahade@nasa.gov}

\author[orcid=0000-0001-6590-3479]{Erika Palmerio} 
\affiliation{Predictive Science Inc., San Diego, CA 92121, USA}
\email{epalmerio@predsci.com}

\author[orcid=0000-0002-7676-9364]{Evangelia Samara} 
\affiliation{Heliophysics Science Division, NASA Goddard Space Flight Center, Greenbelt, MD 20770, USA}
\affiliation{The Catholic University of America, Washington, DC 20064, USA}
\email{evangelia.samara@nasa.gov}

\author[orcid=0000-0003-1377-6353]{Phillip Hess}
\affiliation{U.S.\ Naval Research Laboratory, Washington, DC 20375, USA}
\email{phillip.n.hess2.civ@us.navy.mil}

\author[orcid=0000-0002-3253-4205]{Robin C.\ Colaninno}
\affiliation{U.S.\ Naval Research Laboratory, Washington, DC 20375, USA}
\email{robin.c.colaninno.civ@us.navy.mil}

\author[orcid=0000-0003-0565-4890]{Teresa Nieves-Chinchilla}
\affiliation{Heliophysics Science Division, NASA Goddard Space Flight Center, Greenbelt, MD 20770, USA}
\email{teresa.nieves@nasa.gov}

\begin{abstract}
On April 21, 2023, a complex filament eruption produced an Earth-directed coronal mass ejection (CME) that triggered the first severe geomagnetic storm of solar cycle 25. The primary objective of this work is to understand the physical origin of the severe impact of this CME on Earth. We explore the CME's complex morphology by simulating the eruption using the thermodynamic magnetohydrodynamic (MHD) model CORHEL-CME. We model the pre-eruptive structure as two intertwined magnetic flux ropes and validate the {pre-eruptive configuration and the resulting CME morphology} by comparing synthetic observables generated from the model with extreme-ultraviolet and white-light observations. We then compare the simulated CME structure with in situ measurements. From our analysis, we conclude that the CME merged with a stream interaction region (SIR) before reaching Earth. {This interaction compressed the western flank of the CME, progressively building up a sustained negative $B_n$ component that, together with the associated enhancement in solar wind density and velocity, drove the storm to severe levels. The Dst index computed from the simulated solar wind reaches the observed minimum of $-213\,$nT, although the early phase of the storm is not reproduced. The steady-state background solar wind shifts the longitude of best agreement by ${\sim}20^\circ$ and limits the comparison with other in situ viewpoints. Nevertheless, this model--data comparison illustrates how MHD simulations constrained by multi-viewpoint observations can reveal} hidden dynamical processes and improve the understanding and prediction of space weather impacts.
\end{abstract}







\section{Introduction} \label{sec:intro}

Coronal mass ejections (CMEs) originate from the expulsion of plasma and magnetic fields from the corona into interplanetary space. They represent one of the most significant drivers of space weather and can produce major disruptions to technological systems and human activities \citep{eastwood_2017}.

Despite decades of study, the evolution and interaction of CMEs with the interplanetary magnetic field, from its origin to 1$\,$au, remains not fully understood for two connected reasons. On the one hand, the magnetic field that governs CME evolution cannot be measured directly with remote sensing instruments. While the density structures observed with coronagraphs and heliospheric imagers are thought to trace the underlying magnetic topology, how faithfully white-light morphology reflects the internal magnetic configuration remains an open question \citep[e.g.,][]{vourlidas_2014, alhaddad_2025}. At the same time, only a handful of spacecraft provide in situ measurements at any given time, leaving most of the CME journey unobserved. This combination of sparse spatial coverage and the observational gap is fundamental to why space weather forecasting remains so difficult \citep{nieves_2013, manchester_2017, nieves_2018, kilpua_2019, luhmann_2020}. {Notably, this difficulty does not diminish with heliocentric distance: while the CME's early dynamics near the Sun are shaped by largely unconstrained Lorentz forces, its later evolution is instead shaped by interactions with the structured solar wind, such as stream interaction regions (SIR) or other CMEs, that are similarly difficult to constrain from either remote-sensing or in situ observations alone, and that operational forecasting models do not typically resolve.}

CMEs erupt from the Sun carrying ordered, twisted magnetic structures known as magnetic flux ropes \citep[MFR;][]{low_1984, chen_1996, vourlidas_2014, patsourakos_2020}, whose orientation and helicity can determine whether they will trigger a geomagnetic storm upon arrival at Earth \citep{dungey_1961, kamide_1998}. However, relating solar and in situ observations is hindered by two limitations: even if a CME were to arrive completely unaltered, it would still be sampled only locally, making it difficult to infer global flux rope properties from a single spacecraft crossing; and beyond that, the structure evolves considerably as it propagates from the Sun to Earth. As a CME propagates through the heliosphere, it continuously interacts with the ambient solar wind and magnetic field, exchanging field lines through reconnection, eroding its outer layers, and responding to local distortions that cannot be extrapolated to the entire structure \citep{ruffenach_2012, owens_2017}. Rather than traveling as a coherent, isolated entity, a CME progressively merges with its surroundings, so that what arrives at Earth is not simply the original ejecta but an evolved structure shaped by the entire journey \citep{demoulin_2010, alhaddad_2019, alhaddad_2025}. More critically for space weather purposes, CMEs undergo rotation and deflection that alter the orientation of the flux rope axis and therefore the sign and magnitude of $B_n$ at Earth \citep{nieves_2012, isavnin_2013, palmerio_2018}. 

Moreover, studies of the most intense geomagnetic storms consistently show that they are mostly associated with complex conditions rather than isolated CMEs \citep[e.g.,][]{temmer_2012, lugaz_2017}. CME--CME and CME--SIR interactions can compress the sheath, enhance magnetic field strength, and sustain southward $B_n$ intervals far beyond what the CME alone would produce \citep[e.g.,][]{scolini_2020, palmerio_2022}. However, such interactions produce complex signatures in remote-sensing observations that are difficult to disentangle or occur outside the FOV of the white-light instruments, and by the time the structure arrives at 1$\,$au, the contributing solar wind components may have fully merged into a single, apparently coherent transient. Understanding when and how these interactions occur, and which of them is responsible for driving a given storm to severity, is therefore one of the central open problems in space weather prediction.

Addressing this challenge requires a framework that can simultaneously constrain what we can observe and reconstruct what we cannot. Multi-viewpoint remote-sensing observations provide the first pillar of this framework. A new generation of missions, such as Solar Orbiter \citep[SolO;][]{muller_2020} and Parker Solar Probe \citep[PSP;][]{fox_2016}, provide both in situ magnetic field measurements and remote-sensing observations from novel vantage points in the inner heliosphere. 
The Solar Orbiter Heliospheric Imager \citep[SoloHI;][]{howard_2020} offers high-resolution images, resolving fine-scale structures in CMEs that were previously undetectable \citep{hess_2023}. Moreover, the orbit of SolO places SoloHI at a significantly different longitude from Earth, enabling side-view tracking of CMEs that complements Earth-based perspectives and reveals complexity invisible from a single viewpoint \citep{maccormack_2025b}. Together with data from the Solar and Heliospheric Observatory \citep[SOHO;][]{domingo_1995} and the Solar Terrestrial Relations Observatory \citep[STEREO;][]{kaiser_2008}, these platforms allow multi-viewpoint analyses that provide constraints on CME morphology and kinematics throughout the heliosphere. Physics-based magnetohydrodynamic (MHD) models provide the second and equally essential pillar. Frameworks such as CORHEL-CME \citep{linker_2024} enable the magnetic configuration of a CME to be reconstructed continuously from eruption to arrival, filling the observational gaps where the magnetic field evolves unseen. By anchoring these simulations to multi-viewpoint observations at both ends of the CME journey, it becomes possible to investigate the hidden portion of CME evolution (i.e., the reconnection and deformation(s) that occur between the corona and the in situ observations) and to disentangle what is intrinsic to the eruption from what is acquired during propagation. 

The power of this combined framework has been demonstrated in two preceding studies. \citet{maccormack_2025} showed that a single eruption can produce a complex white-light morphology, with an apparent structure shaped by early interactions between the flux rope and the ambient coronal magnetic field. Building on this, \citet{sahade_2026} showed that even modest variations in pre-eruptive magnetic configurations lead to significantly different CME trajectories and morphologies, underscoring how sensitively CME evolution depends on initial conditions and how necessary data-constrained modeling is to interpret what eventually arrives at 1$\,$au. Together, these works establish that reproducing observed CME complexity requires both a well-constrained initial magnetic configuration and a realistic coronal and heliospheric environment, and that the model, when properly anchored to observations, can indicate the underlying magnetic structure and guide the interpretation of the observed CME signatures.

In this work, we apply this combined framework to the first severe geomagnetic storm of Solar Cycle 25, which occurred in April 2023 \citep{hajra_2024, vemareddy_2024, paouris_2025, kumar_2026, lugaz_2026, sarkar_2026}. 
{The main motivation is that the severity of the event was unexpected: NOAA's operational forecasts evolved substantially over the days leading up to the event, illustrating the difficulty of predicting this CME's impact from remote-sensing observations alone. On April 22, with the CME identified as an Earth-directed event, the forecast stated no G1 or greater storm was expected\footnote{\href{https://www.ngdc.noaa.gov/stp/space-weather/swpc-products/daily_reports/3day_forecast/2023/04/202304220030three_day_forecast.txt}{NOAA three-day forecast, April 22, 2023}}. On April 23, hours before the CME shock arrived at Wind, the forecast was raised to G1--G2 (Kp up to 6.33)\footnote{\href{https://www.ngdc.noaa.gov/stp/space-weather/swpc-products/daily_reports/3day_forecast/2023/04/202304230030three_day_forecast.txt}{NOAA three-day forecast, April 23, 2023}}. The storm that ultimately developed reached Kp\,=\,8\footnote{\href{https://doi.org/10.5880/Kp.0001}{GFZ Potsdam Data Services}} (NOAA Scale G4, severe), exceeding even the most updated pre-arrival forecast by two storm categories. This under-prediction, despite the CME being tracked in real time, is the specific, quantitative basis for describing the event as unexpected, independent of the CME's speed or source region.}

From Earth's perspective, the associated moderated-speed CME appeared as a faint halo, offering little diagnostic information on its magnetic structure. Solar Orbiter, located 125$^{\circ}$ in longitude away from Earth, provided a complementary side view with SoloHI, revealing a far more complex and structured CME. We use this multi-view geometry, combined with MHD modeling, to reconstruct the magnetic evolution of the CME. {We validate the pre-eruptive configuration and the resulting CME morphology against multi-viewpoint} extreme ultraviolet (EUV) and white-light observations using synthetic images as directly comparable observables, then we follow the CME evolution to 1$\,$au. The central question that we aim to address is whether the unexpected severity of the April 21 event is encoded in the CME's observable solar structure, or whether it emerges from the hidden heliospheric evolution that only a physics-based model can reveal, in the absence of relevant observational constraints. 

In Section~\ref{sec:data} we describe the observations of the event from its source region up to 1$\,$au. In Section~\ref{sec:models} {we model the eruption and early evolution of the CME using the MHD simulation, generating synthetic EUV and white-light observables that we compare against the available observations.} The in situ observations are compared and discussed with the model in Section~\ref{sec:IS}. Finally, we summarize our conclusions in Section~\ref{sec:conc}.

\section{Observations} \label{sec:data}

On 2023 April 23, the first severe geomagnetic storm of solar cycle 25 occurred. The event was driven by an Earth-directed CME that erupted on April 21. To track the evolution of this eruption from its solar origin to its geoeffective consequences, we combine remote-sensing observations of the source region, corona, and heliosphere with in situ measurements near 1$\,$au.

The CME originated from the magnetic field surrounding active region (AR) 13283. The associated filament was located between AR~13283 and its dispersed positive polarity along an extended polarity inversion line (PIL). A coronal hole was located to the south of the eruptive region and may have influenced the trajectory of the CME \citep{moestl_2015, paouris_2025}. 

\begin{figure*}[ht!]
\epsscale{1.1}
\plotone{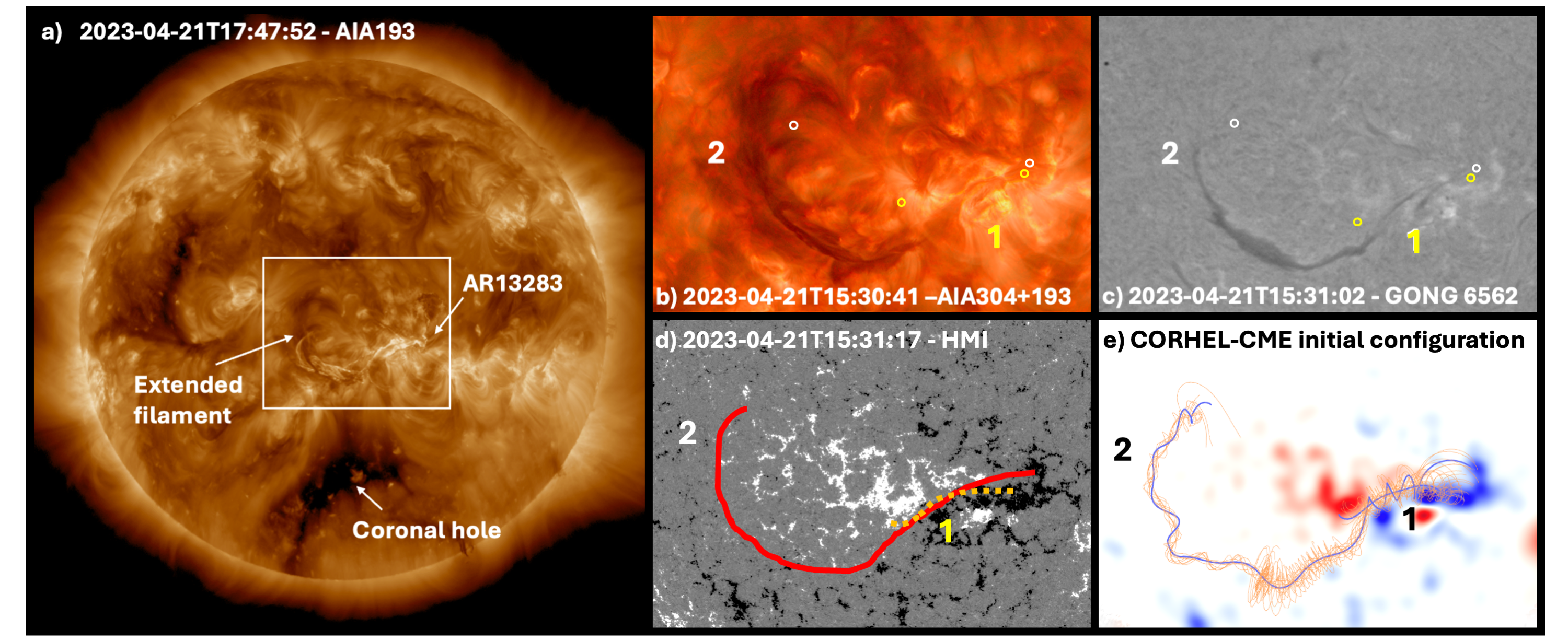}
\caption{{Panel a) Full disk SDO/AIA 193$\,${\AA} observations on April 21 at 17:47$\,$UT. White square denotes the region of the eruption. AR~13283, the extended filament, and a southern coronal hole are indicated with white arrows. Right panels: pre-eruptive configuration observed by SDO/AIA 304+193$\,${\AA} (b) and GONG H-alpha (c) at $\sim$15:31$\,$UT. The extended filament channel (2) is clearly visible in the AIA 304+193$\,${\AA} image, while the eruptive portion (1) is highlighted in the GONG H-alpha image. White and yellow circles schematically represent the footpoints of the long filament (white) and the eruptive portion (yellow) in panels b and c. Panel d) schematically shows the location of the extended filament channel (solid red line, marked 2) and the eruptive portion (dashed yellow line, marked 1) overlaid on the SDO/HMI magnetogram at 15:31$\,$UT. Panel e) shows the corresponding initial condition of the two interacting magnetic flux ropes used in the CORHEL-CME simulation, overlaid on the HMI photospheric magnetogram, with footpoints of the long (2) and eruptive (1) flux ropes indicated.} \label{fig:Source_0421}}
\end{figure*}

{The pre-eruptive configuration, eruption, and subsequent post-eruptive magnetic reconfiguration can be tracked across multiple wavelengths. Figure~\ref{fig:Source_0421} illustrates the pre-eruptive configuration together with the corresponding flux rope geometry adopted in this work (Section~\ref{sec:models}). Panel a shows a 193$\,${\AA} full-disk image, using the Atmospheric Imaging Assembly \citep[AIA;][]{lemen_2012} on board the Solar Dynamics Observatory \citep[SDO;][]{pesnell_2012}, contextualizing the location of AR~13283, the extended filament, and the southern coronal hole. Panel b shows composite AIA 304+193$\,${\AA} imagery from the region of interest, highlighting the extended filament channel prior to eruption. White and yellow circles in panels b–c mark the footpoints of the extended filament (2) and the eruptive portion (1), respectively. Panel c shows H-alpha observations from the Global Oscillation Network Group \citep[GONG;][]{harvey_1996}, capturing the filament before eruption. Panel d shows a schematic representation, based on the EUV and H-alpha observations, of the filament configuration plotted over the line-of-sight magnetic field from the Helioseismic and Magnetic Imager \citep[HMI;][]{scherrer_2012} on board SDO. Panel e shows the corresponding initial condition of the two interacting MFRs used in the CORHEL-CME simulation (Section~\ref{sec:models}) overlaid on the HMI photospheric magnetogram.}

{The filament channel (red solid line, marked '2' in Figure~\ref{fig:Source_0421}d) was notably long and appears to consist of two distinct helical segments. It also highlights the western portion with a dashed yellow line ('1'). The observations suggest that this segment, located along the AR~13283 PIL, initiated reconnection in the vicinity of a magnetic null point and produced the main eruption. The eruption began at approximately 17:34$\,$UT. During the initial phase, the filament western portion ('1') reconnected with the larger filament channel, producing two important effects: first, the erupting structure drags material from the extended filament channel, forcing a reconfiguration within a dispersed region of positive polarity; {second, material from the short filament is pulled by the remaining channel. After this, no multi-viewpoint EUV coverage is available to directly track this early evolution, as STEREO-A was positioned near the Sun–Earth line at the time, and Solar Orbiter’s vantage point could not resolve the low-coronal evolution. Nevertheless, a tilt discrepancy is later observed between the {orientation derived} from the AR and the one reconstructed with the white-light coronagraph images of the associated CME \citep[][]{vemareddy_2024,paouris_2025}.}

\begin{figure*}[ht!]
\epsscale{1.15}
\plotone{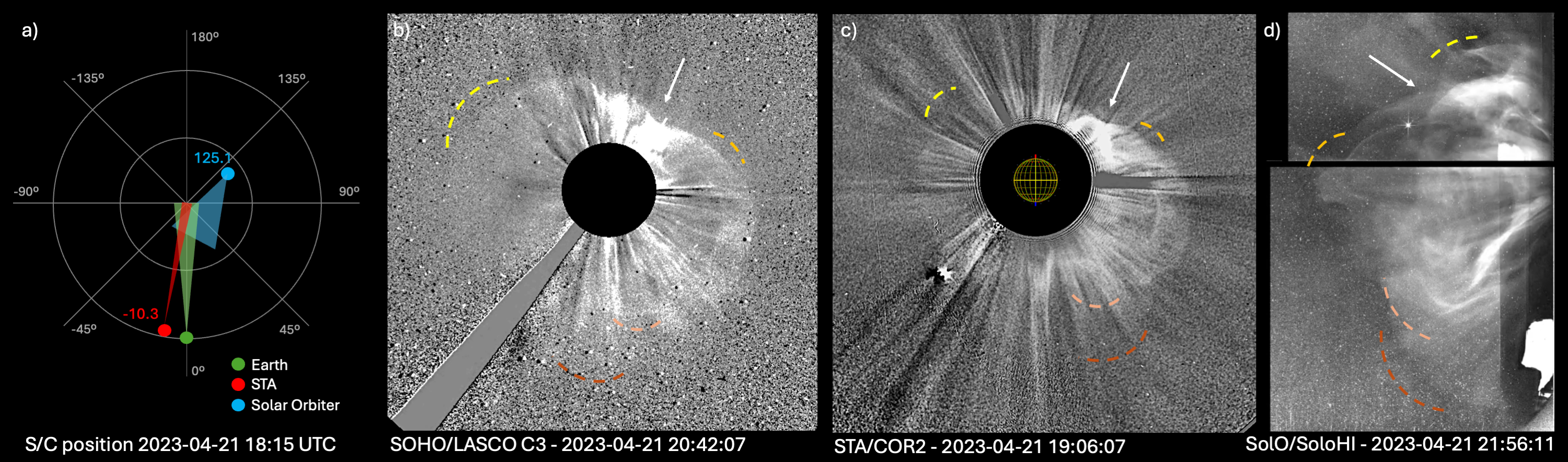}
\caption{Panel a) Spacecraft configuration on April 21, 2023. Panel b) LASCO C3 base difference image for the April 21 event at 20:42$\,$UT. Panel c) COR2 image processed with base difference at 19:06$\,$UT. Panel d) SoloHI inner FOV observations with a background removal process at 21:56$\,$UT. In all panels, dashed curves denote features mentioned in the main text.
\label{fig:RS_0421}}
\end{figure*}

Following the eruption, the CME was extensively observed in white-light imaging from multiple vantage points. The Large Angle and Spectrometric Coronagraph \citep[LASCO;][]{brueckner_1995} on board SOHO, the COR2 coronagraph part of the Sun-Earth Connection Coronal and Heliospheric Investigation \citep[SECCHI;][]{howard_2008} instrument suite onboard the STEREO-A spacecraft, and SoloHI on board SolO provide complementary views of the expanding structure.
On April 21, SolO was positioned at 0.38$\,$au and 125$^{\circ}$ longitude from Earth, while STEREO-A was located at $-10^{\circ}$ relative to Earth (Figure \ref{fig:RS_0421}a). This configuration effectively produced a two-viewpoint scenario since the perspectives from STEREO-A and SOHO reveal very similar large-scale CME morphologies (Figure~\ref{fig:RS_0421} panels b and c).

From the STEREO-A and SOHO viewpoints, the CME is visible on the right side of the image, with a broader shock front extending to the southwest (maroon dashed line). The main body of the CME is clearly identifiable, accompanied by the extended shock feature (brown dashed line, Figures~\ref{fig:RS_0421}b and c) and a density enhancement to the north (white arrow). The eastern side of the CME (yellow dashed line) appears less dense than the western side (orange and peach dashed lines), although both remain part of the main CME structure. From this perspective, the CME appeared as a faint halo, and previous analyses place its speed in the moderate range \citep[$1000-1500$ km s$^{-1}$,][]{paouris_2025, lugaz_2026}. As mentioned before, with these observational constraints, space weather forecasters predicted only a minor geomagnetic impact for this event, reflecting an incomplete picture of the CME's geoeffective potential. This suggests that additional factors not captured by the available observations or operational models played a role in shaping the storm's severity, making this event a compelling case for modeling.

In contrast, the SoloHI observations (Figure~\ref{fig:RS_0421}d), which were not available in real time due to data latency and therefore not used in the original forecast, provides a side-on perspective that reveals a significantly more complex and structured CME morphology. Several distinct fronts can be identified from this viewpoint. Based on the latitude of the fronts, they can plausibly match those seen in the coronagraphs.  A narrow-shaped front (orange dashed line) may correspond to the dense core of the CME observed in the coronagraph images, while an extended lower front (brown dashed line) in perspective appears to propagate faster than the northern structures. Additionally, a pronounced density enhancement (white arrow) is visible in the northern region, possibly resulting from the overlap of multiple fronts. A V-shaped structure at the back of the CME may indicate the presence of distinct plasma components within the ejecta that would be partially obscured by the occulter in the COR2 and C3.\\

As the CME propagated through the heliosphere, it was eventually detected in situ by spacecraft near 1$\,$au. The CME impacted Wind \citep{ogilvie_1997} and STEREO-A on April 23 at approximately {16:59$\,$UT} and {14:31$\,$UT}, respectively. Figure \ref{fig:IS_0421} shows the magnetic and plasma observations from both spacecraft between April 23 and April 26. Panel (a) and (b) display the total magnetic field and its RTN components, respectively. Panels (c) and (d) show the solar wind velocity as well as the density and temperature. The Dst index\footnote{\href{http://wdc.kugi.kyoto-u.ac.jp/dstdir/}{World Data Center for Geomagnetism, Kyoto}} is presented in panel (e).

\begin{figure*}[ht!]
\epsscale{1}
\plotone{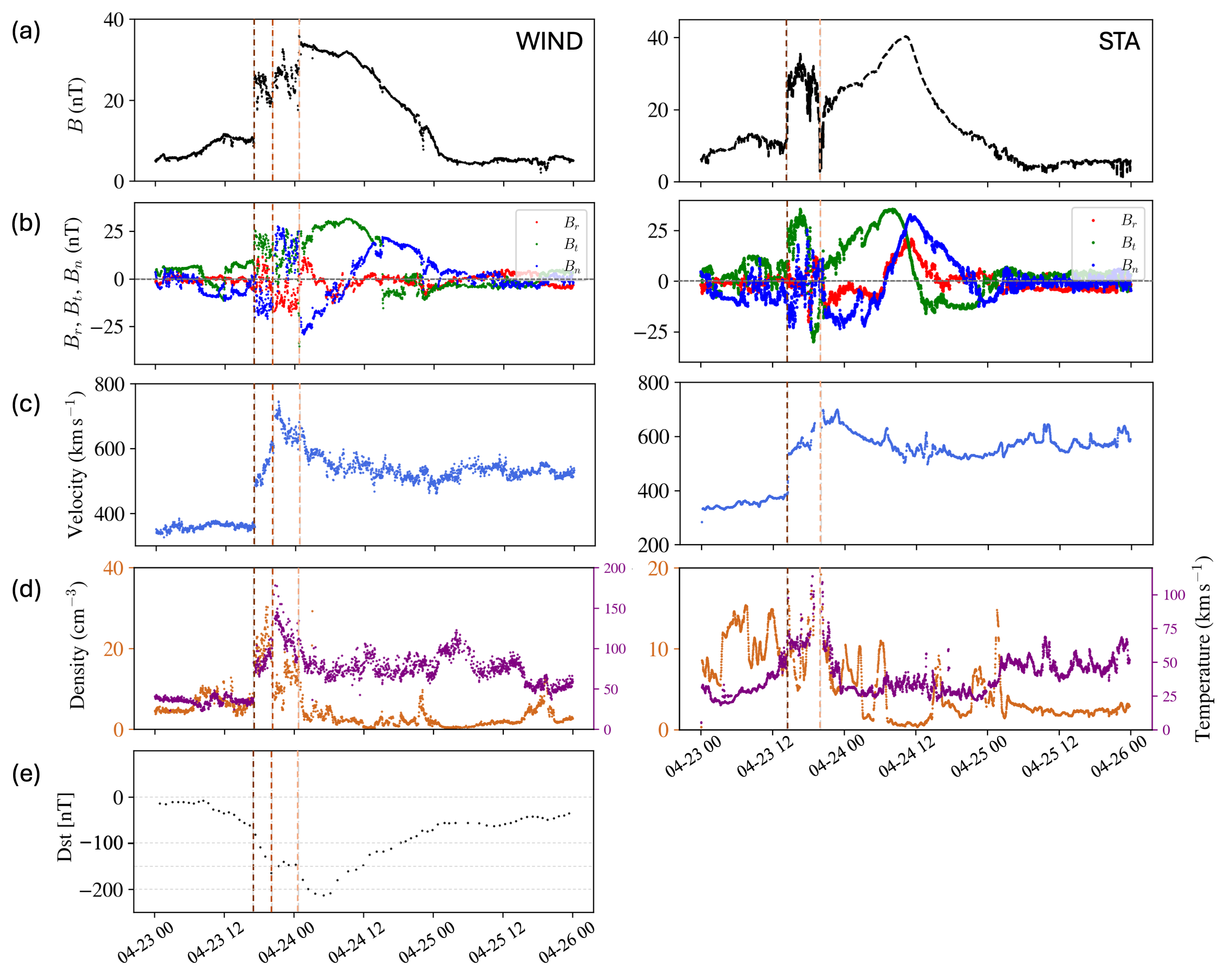}
\caption{In situ measurements for WIND (left) and STA (right) between April 22 and April 27. Panels a and b show the total magnetic field and its components; in the components panel, a horizontal gray dashed line at 0 is included for reference. Panels c and d show the velocity and density and temperature, respectively. Panel e shows the Dst index only for WIND spacecraft. Vertical lines denote the shock arrival and the start of the MO for both spacecraft. \label{fig:IS_0421}}
\end{figure*}

At Wind, the arrival of the CME is marked by a shock at {16:59$\,$UT} on April 23 (brown vertical lines in Figure \ref{fig:IS_0421} left). The sheath region exhibits a sustained negative $B_{n}$ component from the shock arrival until approximately {20:16$\,$UT} on April 23 (maroon vertical lines in Figure \ref{fig:IS_0421} left), after which changes in plasma parameters (density and temperature) and magnetic field components indicate a transition within the sheath, suggesting the presence of an additional embedded structure. Perturbations in the background solar wind coincide with the initial decrease in the Dst index while the shock arrival coincides with the onset of an intense geomagnetic storm \citep[Dst $<-$100;][]{gonzalez_1994}. The later sheath interval (maroon vertical lines in Figure \ref{fig:IS_0421} left) is temporally associated with the beginning of a short recovery phase. The magnetic obstacle (MO), defined here as the interval of smoothly rotating magnetic field components and depressed temperature indicative of the ejecta interior, starts at {00:58$\,$UT} on April 24 (orange vertical lines in Figure \ref{fig:IS_0421} left).

A second decrease in Dst occurs at the onset of the MO, where the $B_{n}$ component becomes negative and remains southward for at least 9 hours. This interval further deepens the storm, driving the Dst index to a minimum of -213$\,$nT at {05:13$\,$UT} on April 24, after which $B_{n}$ gradually turns northward and becomes positive after {09:45$\,$UT}. 

{At STEREO-A, the shock arrived earlier, at around 14:28$\,$UT on April 23 (brown vertical lines in Figure \ref{fig:IS_0421} right). It also exhibits a small period of southward $B_{n}$ component within the sheath, although it is notably shorter than that observed at Wind. In contrast to Wind observations, no clear substructures can be identified within the sheath, which instead presents smoother variations without clear embedded substructures until the MO appears at 20:29$\,$UT (orange vertical lines in Figure \ref{fig:IS_0421} right). During this interval, the $B_{n}$ component remains negative for approximately 10.5 hours.}

{For a more complete description of the in situ signatures, the reader can refer to \cite{lugaz_2026}. After an analysis of the CME using the combined observations from Wind and STEREO-A, their results indicate that the sheath consists of an initial interval of compressed solar wind followed by a more variable region with draping-like magnetic fields and marked differences between spacecraft, highlighting strong spatial variability and a lack of a simple relationship between spacecraft separation and measurement coherence. Overall, these results emphasize the limitations of single-point observations and suggest that commonly assumed relationships between CME geometry and in situ signatures may not hold for individual events.

Understanding how the April 21 CME ultimately drove a severe geomagnetic storm requires a modeling framework capable of connecting observations across all stages of its evolution. In particular, the relationship between the complex solar eruption, the halo observed from Earth and the highly structured morphology observed by SoloHI, as well as its relevance to the in situ measurements can be explored by modeling the CME evolution continuously from its solar source to 1$\,$au. Data-inspired MHD simulations provide a means to bridge these observational gaps and test whether a unified physical description can reproduce the multi-spacecraft signatures. In the following section, we employ the CORHEL-CME MHD model to reproduce the event and attempt to reconcile and explain the remote-sensing and in situ observations within a single physical framework.}

\section{Analysis}
\label{sec:models}

CORHEL-CME is a modeling suite that enables self-consistent, physics-based simulations of CME eruption and propagation from 1\,$R_{\odot}$ to 1\,au and beyond. It is based on the Magnetohydrodynamic Algorithm Outside a Sphere \citep[MAS;][]{mikic_2018, downs_2025} code, which solves the time-dependent, resistive, thermodynamic MHD equations on a non-uniform spherical mesh. CORHEL-CME features two coupled spatial domains: COR, which describes the corona in the range 1--30\,$R_{\odot}$, and HEL, which describes the heliosphere in the range 28--236\,$R_{\odot}$. COR is initialized from a global map of the photospheric magnetic field and employs a heating model to produce a fully thermodynamic MHD corona, while HEL is driven by the COR solution at 28\,$R_{\odot}$. Modeling the eruption and propagation of CMEs within this framework involves constructing an eruptive magnetic flux rope (MFR) rooted in the photosphere along a PIL, described using the highly flexible Regularized Biot--Savart Law \citep[RBSL;][]{titov_2018} formulation. For additional technical details on the model, the reader is referred to \citet{linker_2024}.

To model the coronal and heliospheric conditions during the 2023 April 21 event, we use magnetograph data from SDO/HMI. Specifically, we combine data from Carrington Rotations (CRs) 2269 and 2270 to construct a magnetogram in which each heliolongitude is represented by observations taken as close as possible to the CME eruption time. We then replace regions within $120^{\circ}$ of the Earth-facing central meridian with data from a magnetogram snapshot taken on the day of the eruption, following the approach used for HMI synchronic daily products \citep{hayashi_2015}. To design the MFR to insert within AR~13283 and its surrounding region, we employ magnetograph data from SDO/HMI and extreme ultra-violet (EUV) observations from SDO/AIA to infer pre-eruptive properties of the CME such as footpoint location, helicity sign, and axial magnetic field direction \citep[guided by established observational diagnostics; see, e.g.,][]{palmerio_2017}. This approach---combining photospheric magnetic field measurements with coronal EUV signatures to inform flux rope construction---has also been employed in previous CORHEL-CME studies \citep[e.g.,][]{maccormack_2025b, barnes_2026, sahade_2026}.

As mentioned in Section~\ref{sec:data}, the western portion of the elongated filament drove the main eruption. However, the event also involved the activation of strands belonging to the entire filament channel, resulting in a complex and distorted PIL structure that cannot be adequately represented by a single initial MFR. We therefore model this event by constructing two partially overlapping MFRs \citep[see][for a similar approach]{torok_2018}{, shown in Figure~\ref{fig:Source_0421}e}. The positions and footpoints of both the elongated section and the eruptive portion are determined using the tie-pointing technique described in \citet{sahade_2024}. Hence, the 3D information on the location and evolution of the filament channel is used both as a guidance to set up the simulation's initial conditions {(Figure~\ref{fig:Source_0421}e)} and as an early validation of the CORHEL-CME simulation. 

The two flux ropes {are built with identical} apex heights (1.025$\,R_{\odot}$)
and {cross-sectional} radii (0.015$\,R_{\odot}$), but differ substantially in their {internal current, with the short, eruptive segment carrying approximately three times the current of the longer, stable filament segment.}
This difference reflects a higher degree of non-potentiality assigned to the short, eruptive segment, consistent with its role as the component that becomes unstable and drives the eruption, while the longer, stable filament segment was constructed with a comparatively lower twist, consistent with its more gradual, dragged evolution during the eruption. The complete set of parameter values for both flux ropes is publicly available through the CCMC Run-on-Request archive\footnote{\href{https://ccmc.gsfc.nasa.gov/ror/results/viewrun.php?runnumber=Abril_Sahade_112525_SH_1}{A. Sahade run 112525\_SH\_1}}, enabling full reproducibility of the initial conditions.

{It is worth noting that this choice of a two-MFR configuration is supported by the agreement with the observations. {We tested three initial configurations: a single short MFR, a single long MFR, and the two interacting MFRs presented here, and found that only the two-MFR configuration reproduced the available observations of the evolution: the pre-eruptive EUV configuration and the later white-light coronagraph morphology. This agreement between the available observations and synthetic observables from the model is consistent with the apparent tilt change reported between AR-based and coronagraph-based reconstructions of this event \citep{vemareddy_2024,paouris_2025}.} We do not show the non-reproducing configurations here, as our focus is on the configuration that best reproduces the observed eruption.}

Figure \ref{fig:SC_CORHEL_0421} displays the pre-eruptive (upper panels) and post-eruptive (lower panels) configurations of the elongated filament channel. The left column shows CORHEL-CME synthetic EUV emission emulating the AIA 193\,{\AA} channel, while the right column shows base-difference AIA 193\,{\AA} observations to highlight the eruptive signatures. In the pre-eruptive configuration, the elongated filament is labeled as `2' and the eruptive portion as `1'. Although the model is not expected to reproduce every observed detail, it is noteworthy that the main characteristics of the eruption are well captured when comparing the synthetic emission with the AIA observations (see Figure~\ref{fig:Source_0421}). In particular, the location of the ribbons, the dimming region on the right side, and the reconnection of some strands (originating from the elongated filament) within the dispersed positive polarity region near the eruptive portion serve as strong indicators of the model's accuracy in capturing the eruptive process. Successful reproduction of these characteristics supports the choice of modeling this eruption as two different MFRs that interact when one portion begins to erupt.

\begin{figure}[ht!]
\epsscale{1.17}
\plotone{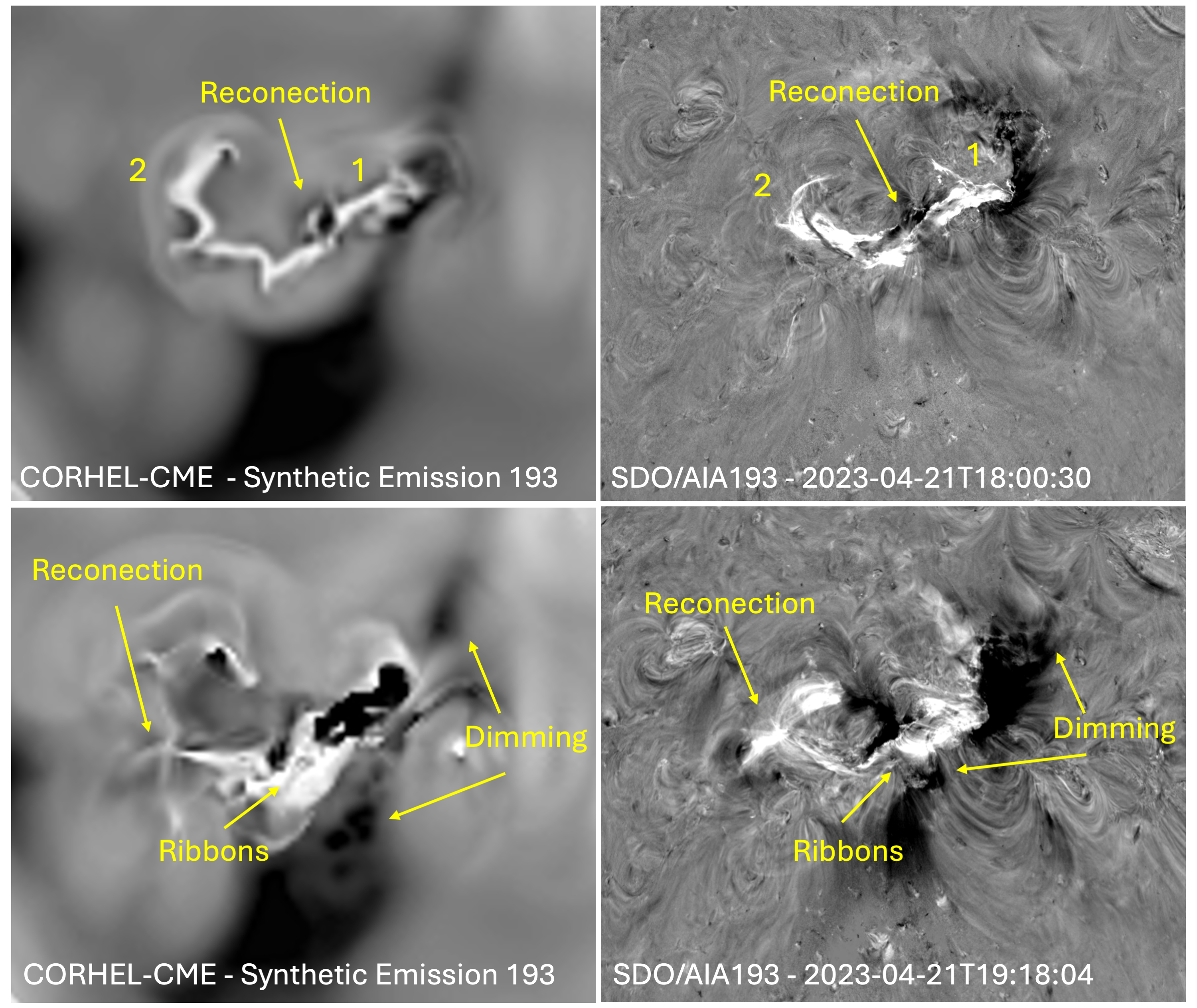}
\caption{Pre (upper panels) and post-eruptive (lower panels) configurations of the CME source region. Left column shows synthetic EUV emission at 193$\,${\AA} from the CORHEL-CME model; right column shows the corresponding SDO/AIA 193$\,${\AA} observations. The elongated filament channel (2), eruptive portion (1) are indicated in the upper panel, while ribbons and dimming regions are marked in the lower panels by yellow arrows. \label{fig:SC_CORHEL_0421}}
\end{figure}

\subsection{White-light structures} \label{WLst}

\begin{figure*}[ht!]
\epsscale{1.15}
\plotone{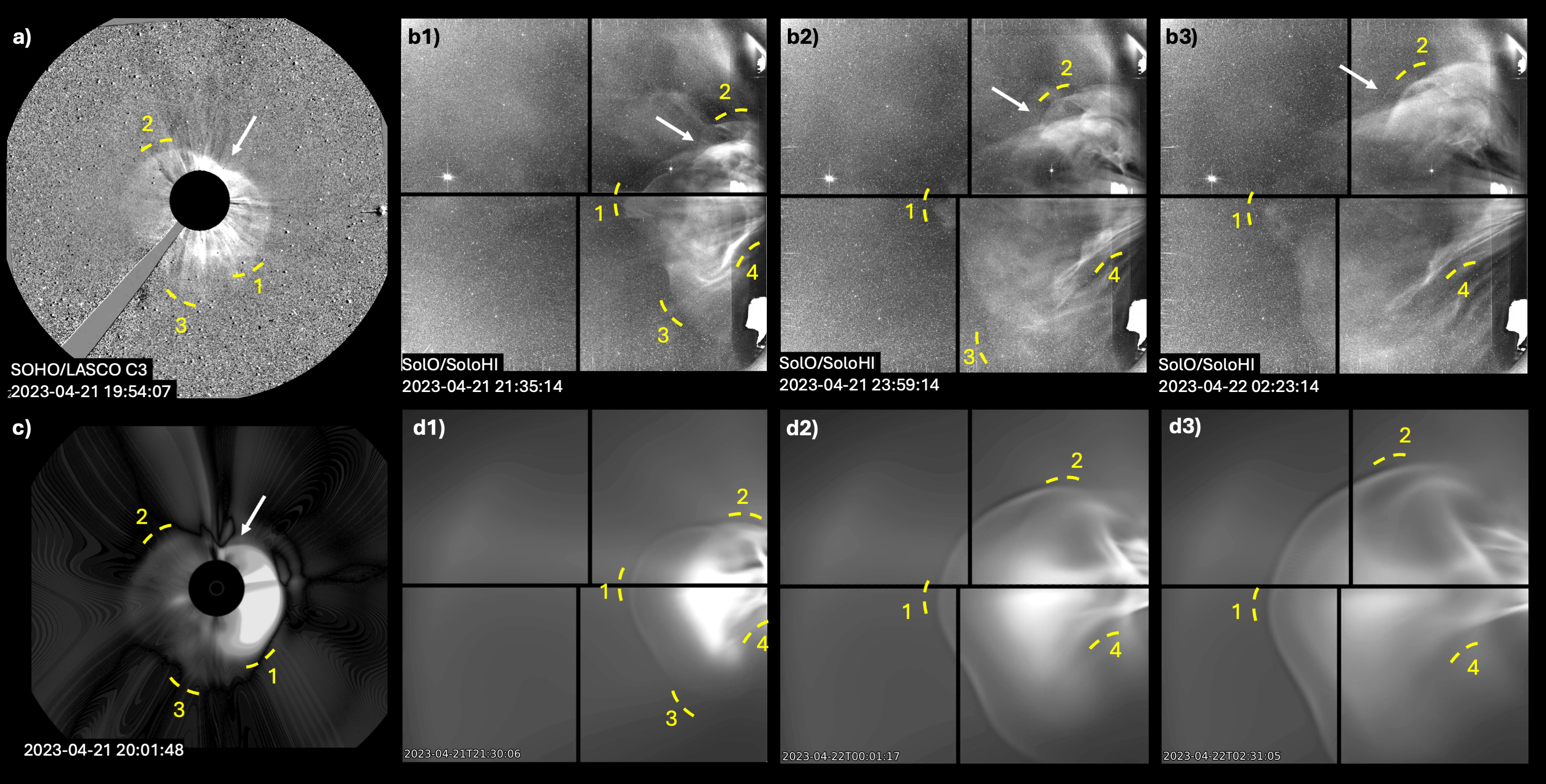}
\caption{Comparison between white-light observations (upper panels) and synthetic images (lower panels). Panel (a) on the left shows base-difference observations of the event from LASCO C3 at 21:54$\,$UT. The remaining panels {(b1-b3)} show the evolution of the CME from the SoloHI point of view at three different times, with a background-removal process applied to the observations. {Panel (c) and (d1-d3) shows the synthetic images obtained from CORHEL-CME results for LASCO C3 and SoloHI FOV, respectively.} All synthetic images are base-difference images with border enhancement applied. White arrows and curves indicate features that can be correlated, as described in the text.\label{fig:WL_CORHEL_0421}}
\end{figure*}

The upper row of Figure~\ref{fig:WL_CORHEL_0421} presents LASCO/C3 and SoloHI observations of the event. The tracked features are indicated with white arrows and yellow curves. Equivalent features are marked in the lower row of the synthetic white-light images generated using the modeled CME. Compared to the preceding studies in this series \citep{maccormack_2025, sahade_2026}, where modeled magnetic field lines were overlaid on white-light images for qualitative comparison, here we generate fully synthetic white-light images from the model output. This represents a more direct and quantitative validation: rather than assessing whether field line morphology is consistent with observed density structures, we compare observables to observables directly.

From the LASCO/C3 perspective, the western part of the  CME is brighter than the eastern part. The main body of the CME, marked as (`1'), is reproduced by the model reasonably well. By tracking the evolution of the magnetic field lines, we identify that the enhanced density region in the upper part, indicated by a white arrow, is where the two modeled pre-eruptive MFRs are interacting. A faint shock structure in the lower part (`3') is also visible in the synthetic images and corresponds to the interaction of the main body of the CME with the southern coronal hole, which leads to its rapid expansion. The eastern portion (`2') appears fainter in both the synthetic and observed white-light images, though it remains identifiable. In the synthetic images, this region shows less accumulated material, consistent with the stable section of the MFR being dragged by the eruptive part without contributing mass. Whether this interpretation extends to the observations is less clear, as the eastern portion appears comparatively diffuse to the rest of the CME structure. The main discrepancy between the model and LASCO C3 observations lies at the northeast front (white arrow), where the model shows more expansion, possibly due to differences in the modeled structure of the coronal magnetic field.

The synthetic SoloHI images reproduce the main observed structural features, although the model is not expected to capture the full complexity of the observations. Throughout the CME's evolution, SoloHI observations provide a favorable viewpoint that compensates for the lack of structure identified in LASCO observations due to the Earth-directed nature of the event. Regions of accumulated density are well defined, and both the conical front of the CME, which correlates with the main body of the CME (`1'), and the extended southern shock structure (`3') in the lower section remain clearly identifiable throughout the evolution. Moreover, the eastern signature seen in the LASCO view can now be correlated with the upper part of the SoloHI CME structure (`2'), which appears to overtake the main body of the MFR (`1') as a projection effect. As a consequence of the side-on viewing geometry of SoloHI, this upper structure extends further along the line of sight, making it appear ahead of the main body of the CME in the plane of the sky. An internal enhancement of density (`4') is also well reproduced and remains coherent throughout the evolution. Altogether, these comparisons validate the choice of initial conditions used for constructing the MFRs and validate the modeling framework to reproduce the complex morphology of the observations.

Importantly, the SoloHI observations reveal a CME that, at this stage of its propagation, has not undergone any significant interactions that would increase its geoeffectiveness. The main fronts are well defined and no strong asymmetry or compression signatures are evident. The absence of such signatures suggests that whatever intensified the geoeffectiveness of this event, if not its initial complexity, happened further out in the heliosphere, beyond the available remote-sensing observations. Although STEREO/HI observations exist for this event, the Earth-directed nature of the CME combined with the near-Earth position of STEREO-A at that time results in an unfavorable viewing geometry, and the limited spatial resolution of HI2 prevents the identification of clear heliospheric signatures. These observations motivate extending the simulation and analysis to 1$\,$au, where the model can access the region not well observed. Building on the agreement between the observations and synthetic images, we analyze the in situ consequences to ultimately answer the central question: How did the April 21 CME result in a severe geomagnetic storm?

\section{Results}
\label{sec:IS}

\begin{figure*}[ht!]
\epsscale{1.10}
\plotone{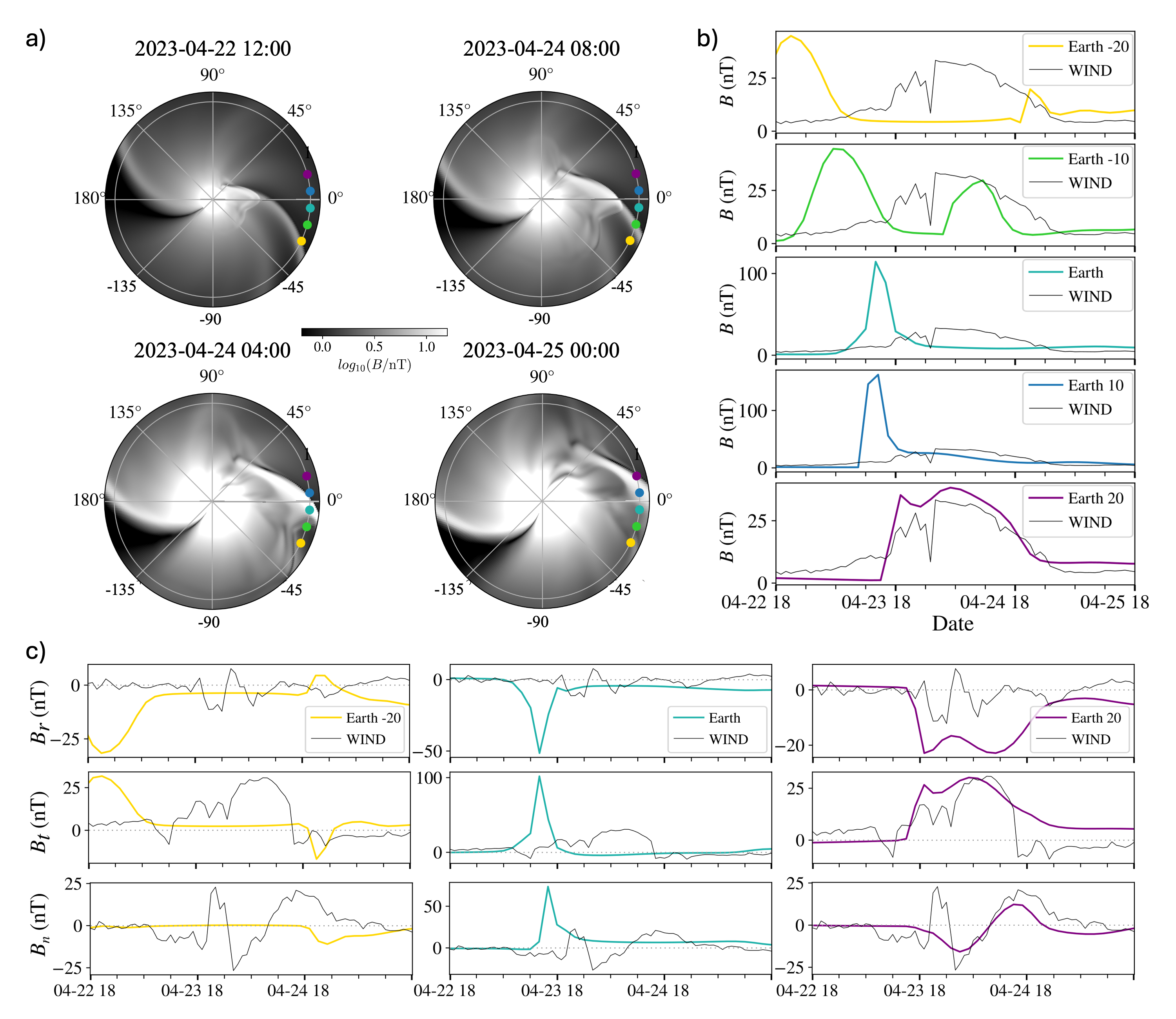}
\caption{Panel a shows slices of the total magnetic field at latitude 6° at different times when the SIR or the CME--SIR structure impacts 1$\,$au. Colored dots represent any of the five representative longitudes from a longitudinal sweep centered on Earth. The full scan is performed in steps of $10^\circ$ toward the east and west. Panel b shows the timeline of the total magnetic field at each one of the five longitudes. Panel c shows the magnetic field components at 1$\,$au for three selected longitudes out of the five shown in panel b (Earth $-20^\circ$, Earth and Earth $+20^\circ$). The black solid line represents the WIND total magnetic field magnitude for comparison (note that scales differ between plots). An animated version of panel a, showing the evolution from {April 21, 18:00$\,$UT to April 25, 18:00$\,$UT}, is available online. \label{fig:CIR-Btot_CORHEL_0421}}
\end{figure*}

As mentioned earlier, the April 21 CME was a moderate-speed \citep[1000$\,$-$\,$1500$\,$km$/$s;][]{paouris_2025} Earth-directed event that, based on its observable properties available in real time, gave no indication of an upcoming severe geomagnetic storm. The severe impact at Earth was primarily driven by a sustained southward magnetic field first in the CME sheath region and later within the MO. In the white-light comparison, the CME showed no signs of significant interaction or structural change within the imager's field of view that would have motivated updating the original forecast, suggesting that what drove the intense geomagnetic storm originated further out in the heliosphere. We therefore extend the simulation to 1$\,$au to investigate what role the structured solar wind played in shaping the geoeffective impact of this event. To do so, we analyze the longitudinal variability of the modeled magnetic field at 1$\,$au by sampling the modeled solution at discrete longitudes around Earth’s position. From this sampling, we select five representative longitudes ($-20^\circ$, $-10^\circ$, $0^\circ$, $10^\circ$, $20^\circ$) that capture the global longitudinal variability of the event.

Figure~\ref{fig:CIR-Btot_CORHEL_0421}a shows snapshots of the total magnetic field of the CME and the background solar wind as the CME evolves towards Earth. An animation showing the evolution of $B$ from April 21 at 18:00~UT to April 26 at 22:00~UT is available in the online version of the article. The light-blue dot represents Earth's position in Carrington coordinates. As can be noticed in the figure and its corresponding animation, the complex morphology of the CME, where the two modeled MFRs interacted, developed toward eastern longitudes relative to Earth producing an asymmetric structure. The western flank of the CME, in contrast, propagated between $0^\circ\,$and$\,+10^\circ$ into a pre-existing stream interaction region (SIR), visible as a region of enhanced magnetic field in the background solar wind. As the CME expands, the SIR progressively compresses its western flank while the eastern side expands freely, resulting in the longitudinal variability discussed in the following sections. This is further confirmed by following the evolution of the magnetic field lines of the MFRs from the corona to the heliosphere\footnote{The coronal evolution of the field lines can be consulted at the \href{https://ccmc.gsfc.nasa.gov/RoR_WWW/SH/CORHEL-CME/2025/Abril_Sahade_112525_SH_1/Abril_Sahade_112525_SH_1_cme_thermocme_report/mapfl_sequence/traces.html}{CCMC} interface.}. 

Figure~\ref{fig:CIR-Btot_CORHEL_0421}b shows the total $B$ field every $10^\circ$ in longitude. The Wind observations are included in all panels to identify which modeled longitude best reproduces the in situ measurements at Earth, accounting for possible offsets in the modeled CME propagation direction. Modeled in situ profiles are shifted by 7.5 hours every $10^\circ$, matching simulated and observed times at Earth, in order to facilitate comparisons between profiles. The colors of the time series in panels b (and c) correspond to the colored dots marking each sampled longitude in panel a. 

At longitudes to the east of Earth ($-20^\circ$ and $-10^\circ$), the SIR arrives ahead of the CME as a distinct structure, and the two signatures remain distinguishable at 1$\,$au. At longitudes around and west of Earth, the SIR and CME are interacting, and their signatures progressively merge into a single complex transient by the time they reach 1$\,$au.
For longitudes around Earth, the interaction between the SIR and the CME becomes increasingly strong. In this region, the modeled magnetic field exhibits amplitudes that are unrealistically large compared to the Wind observations, indicating that the combined SIR--CME signal becomes strongly amplified, losing any recognizable structure in the process.

The modeled profiles recover more realistic magnetic field magnitudes. In particular, at $+20^\circ$ west of Earth, the SIR signature is sufficiently moderate to allow a physically plausible interaction with the CME, and the model best reproduces the total magnetic field profile observed at Wind, as shown in the bottom panel of Figure~\ref{fig:CIR-Btot_CORHEL_0421}b. The offset between Earth's position and the longitude of best agreement likely reflects errors in the modeled CME propagation direction, arising from uncertainties in the background solar wind modeling, such as inaccuracies in the global photospheric maps used to drive the model, the absence of far-side information, and limitations associated with the steady-state approach adopted here. These factors can affect the heliospheric comparison even when the early coronal evolution is well constrained by observations, as shown in {\citet{sahade_2026}}. In this case, inaccuracies in the modeled SIR magnetic field strength and timing likely contribute to the offset observed near Earth's longitude, consistent with previous forecast error analyses showing that time-dependent solar wind modeling improves the representation of evolving heliospheric structures \citep[e.g.,][]{samara_2025,lionello_2026}.

{We also note that our EUV and white-light validation (Section~\ref{WLst}) constrains the MFR's morphology and evolution, but does not independently constrain the CME's absolute propagation longitude to the same precision; some contribution to the observed offset from the coronal reconstruction itself cannot be fully excluded.
However, our identification of this mechanism does not rely on the numerical precision of the best-fit longitude alone. At longitudes east of Earth, the SIR arrives as a distinguishable structure ahead of the CME, while at longitudes to the west the two signatures progressively merge, consistent with increasing compression. This systematic pattern, rather than an exact quantitative match at a single longitude, is what supports the proposed mechanism.}

{This same limitation applies to STA that was located at approximately $-10^\circ$ from Earth at the time of the event, which corresponds to approximately Earth$+10^\circ$ in the model's longitudinal frame once the offset described above is taken into account. This longitude falls within the region of strong SIR--CME interaction where the modeled field amplitudes are unrealistically large compared to in situ observations. For this reason, we do not include a quantitative comparison with the STA in situ measurements at this longitude.}

Figure~\ref{fig:CIR-Btot_CORHEL_0421}c presents the comparison of the modeled magnetic field components and the in situ observations across selected longitudes. At Earth $-20^\circ$ (left panels), we can see the effect of the complexity of the CME without the disturbance of the SIR. This longitude shows a moderately negative $B_n$ component but the rotation seen in the in situ data is not reproduced by the model as well as the other components of the magnetic field. The middle panels show the magnetic components at Earth's location. As we mentioned before, the steady-state SIR produced a strong peak in the components that masks any CME structure. Finally, the right panels show the magnetic field components at Earth $+20^\circ$, displaying the result of the merging process between the CME and SIR. At this longitude the model {reproduces the main features of the Wind magnetic field observations}: the sustained negative $B_n$, the rotation of the MO, and the overall profile of the three components are well captured. However, the three-column comparison does more than demonstrate that one longitude provides a better fit than another; it shows that SIR compression on the western CME flank is the essential ingredient required to reproduce the observed in situ measurements. {It is worth noting that while the SIR interaction substantially alters the CME orientation and the resulting $B_n$ profile, the sign of the MFR magnetic helicity is not expected to change through this interaction. Reconnection and compression at the CME–SIR interface primarily redistribute and erode the magnetic field geometry, while magnetic helicity is approximately conserved {during heliospheric propagation} \citep{dasso_2006}.}

\begin{figure}
\epsscale{1}
\plotone{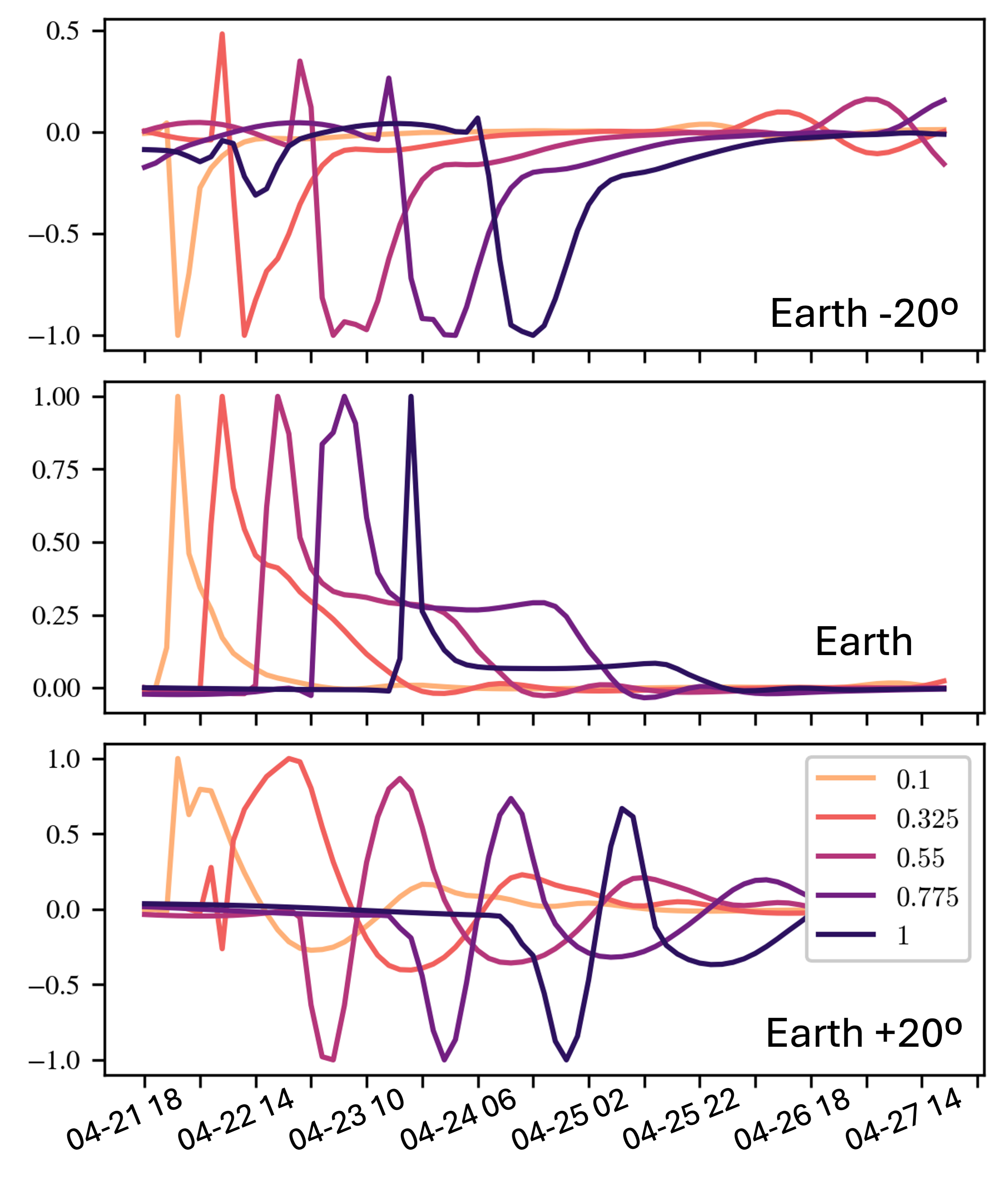}
\caption{Normalized $B_n$ component sampled at three representative longitudes (Earth $-20^\circ$, Earth, and Earth $+20^\circ$) across three heliocentric distances from 0.1 to 1$\,$au, as a function of time since the eruption onset. The contrast between longitudes illustrates the role of SIR compression in building up the southward $B_n$ component during heliospheric propagation. \label{fig:Bz_evolution}}
\end{figure}

To investigate how the SIR interaction builds up the southern component at $+20^\circ$ longitude, we explore how the $B_n$ component evolves with heliocentric distance across three representative longitudes. Figure~\ref{fig:Bz_evolution} shows the normalized $B_n$ at Earth $-20^\circ$, Earth, and Earth $+20^\circ$ across five heliocentric distances from 0.1 to 1$\,$au. {Here, $B_n$ is normalized by its own peak absolute value at each heliocentric distance}. {At Earth $+20^\circ$ (bottom panel), the SIR progressively compresses the western flank of the CME, sustaining the negative $B_n$ and increasing its dominance relative to the positive value with distance, starting at $\sim$0.55\,au, beyond the SoloHI field of view.} By 1$\,$au the structure presents the clear rotation from negative to positive and a $\sim$9$\,$hr of sustained negative $B_n$ as observed in Wind data {(Figure~\ref{fig:IS_0421})}. The contrast with the eastern longitudes is instructive: without SIR compression, the CME expands freely, the negative $B_n$ presents no rotation to positive values, leading to discrepancies with the in situ observations.

\begin{figure}
\epsscale{0.9}
\plotone{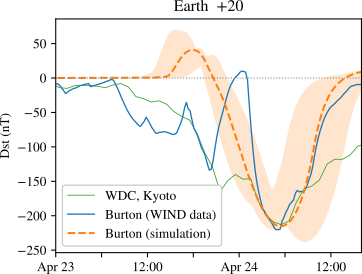}
\caption{{Dst index at Earth$+20^\circ$, computed using the Burton model. The dashed orange line shows the Dst driven by the simulated plasma and magnetic field parameters at this longitude, with the shaded band indicating the uncertainty range obtained by sampling the simulation $\pm2^\circ$ in longitude and latitude. The solid blue line shows the Dst driven by Wind plasma observations (also Burton model), and the solid green line shows the observed Dst index (WDC Kyoto).} 
\label{fig:Dst_Earth20}}
\end{figure}

{The reproduction of the magnetic components and particularly southward $B_n$ excursion at Earth$+20^\circ$ gives us confidence that this is the appropriate longitude to evaluate against the observed storm. As a further validation, we computed the Dst index using the Burton model \citep{burton_1975}. {The Burton model provides an empirical relationship between the interplanetary conditions and the Dst. The empirical Dst is calculated with the magnetic field, density and solar wind velocity. Figure~\ref{fig:Dst_Earth20} shows the comparison between the Dst from the simulation (dashed orange line)}, the Dst index derived from Wind plasma and magnetic field observations using the same Burton model (blue line) and the observed Dst index (WDC Kyoto, green line). {The simulated Dst reaches a minimum comparable to the observed $-213\,$nT, although it does not reproduce the early phase of the storm (see below). This indicates that the combination of magnetic field, density, and velocity captured by the model at Earth$+20^\circ$ is sufficient to produce severe storm levels.} 


As shown in Figure~\ref{fig:CIR-Btot_CORHEL_0421}c, the simulated $B_n$ component fails to reproduce two aspects of the observed profile prior to the start of the magnetic obstacle that are consequential to the observed Dst values. First, the measured background solar wind magnetic field, $B_n$, is approximately $-10\,$nT ahead of the CME's arrival; although small, this negative magnetic field is already sufficient to drive the Dst to approximately $-50\,$nT before the shock arrives. This decrease in Dst ahead of the CME is not captured in our simulation, since it does not reproduce this background southward component of the solar wind. Second, the $B_n$ component of the magnetic field of the observed sheath at Wind changes sign (from negative to positive) at the point of interaction with the SIR, whereas in the simulation $B_n$ is positive before arrival and only turns negative as the CME--SIR system arrives. This reversal is the point at which the simulated Dst begins to decrease; from there, once the simulated $B_n$ matches the magnetic obstacle behavior (along with the other input parameters) the simulated Dst matches the observed minimum Dst.

Relatedly, the model does not capture the sheath region responsible for the storm onset, as the 2$\,$hr output resolution does not resolve shocks or compression structures outside the MFR. Despite this limitation, the simulations reproduce the key feature that intensifies the geomagnetic storm: a sustained negative $B_n$ generated not by the intrinsic properties of the CME, but by its compression against the SIR during heliospheric propagation{, as shown by Figure~\ref{fig:Bz_evolution}, and {supported} by the Dst comparison above (Figure~\ref{fig:Dst_Earth20}).}

{More broadly, our results emphasize the critical role of numerical modeling in space weather forecast error analysis. These simulations reveal the presence of pre-existing large-scale solar wind structures, such as SIRs, that are not clearly identifiable in remote-sensing or in situ observations. Such ``hidden'' interactions can significantly modify subtle properties and enhance geomagnetic activity.}

\section{Discussion and Conclusions}
\label{sec:conc}

On April 21, 2023, a moderate-speed CME produced the first severe geomagnetic storm of Solar Cycle 25, an outcome not anticipated by operational forecasts. Through the combined use of multi-viewpoint observations and CORHEL-CME model, we have {followed the evolution} of this event from its solar source to 1$\,$au and identified {a mechanism that can explain} its unexpected severity: a hidden interaction with a pre-existing SIR, which, although detectable in background solar wind models, was not clearly discernible in the spacecraft data that featured instead a ``clean,'' isolated CME ahead of a solar wind high-speed stream. This interaction compressed the western flank of the CME during heliospheric propagation, sustaining the southward $B_n$ component that{, combined with the enhancement of dynamic pressure, contributed to driving the storm to severe levels.}

{In this work, we trace the eruptive filament in order to select realistic initial conditions to model the evolution of the eruption. We validated the pre-eruptive configuration through comparison with EUV observations and also obtained agreement between the observed complex white-light morphology and the morphology reproduced by the model through synthetic images, further supporting this choice and giving us confidence in the heliospheric solution that revealed the SIR.}

This work represents a methodological step forward relative to the preceding studies in this series. In \citet{maccormack_2025} and \citet{sahade_2026}, modeled magnetic field lines were overlaid on white-light images for qualitative comparison. Here, fully synthetic white-light images allow a direct observable-to-observable validation at each stage of the CME evolution, making the connection between the magnetic structure and what we actually observe more explicit and rigorous. {The $B_n$ evolution across heliocentric distances (Figure~\ref{fig:Bz_evolution}) further demonstrates that the southward component becomes progressively more dominant as the CME propagates (reflected in the shrinking positive peak relative to the sustained negative minimum) indicating that this asymmetry is not an intrinsic property of the CME near the Sun, but builds up progressively through compression by the SIR.}

In addition to the synthetic imagery, we carried out a comparison between the modeled arrival of the CME and the actual in situ measurements taken by the Wind spacecraft. The SIR was not discernible in either the remote-sensing or in situ observations. By the time the merged structure reached Earth, it presented as a single, apparently coherent transient, with no isolated SIR signature in the spacecraft data. {This interaction may be reflected in the observed sheath structure, which could contain remnants of the original SIR material, although a definitive separation of the two contributions is not possible to obtain from in situ data alone.} Moreover, the transition from slow to fast solar wind after the MO passage supports the SIR scenario (as seen in Figure \ref{fig:IS_0421}c). Notably, STEREO-A, located near $\sim-10^\circ$ from Earth, also shows no clear isolated SIR signature, suggesting that the SIR had already merged with the CME by the time the structures reached 1$\,$au. In the model, in contrast, the structures remain separated at 1$\,$au up to ${\sim}-5^\circ$, and interaction occurs at western longitudes. The resulting combined structure presented a prolonged period of southward $B_{n}$ that was a key factor in the severe geomagnetic storm observed at Earth. This model--data comparison {supports the interpretation} that the SIR played a crucial role in shaping the CME's geoeffectiveness, amplifying its impact beyond what would have been expected from the CME alone.

A useful comparison can be made with the August 2018 event, one of the strongest storms of solar cycle 24 \citep[e.g.,][]{nitta_2021, gopalswamy_2022}, in which a SIR interacted with a CME from behind, allowing interaction signatures to remain partially distinguishable in the rear portion of the ejecta. In contrast, for the present event, the CME appears to overrun a pre-existing SIR, leading to its incorporation into the sheath region, where strong compression and turbulence likely obscure its original signatures. These contrasting cases highlight that CME--SIR interactions can produce qualitatively different in situ signatures depending on their relative timing and geometry, challenging forecasting.

{These results should be interpreted within the observational and modeling constraints of this event. The early eruptive phase, between the pre-eruptive EUV observations and the white-light CME, was not captured from independent viewpoints. The model reproduces the orientation at both stages, but how the change in orientation was achieved, through rotation of the flux rope, reconfiguration of the two interacting flux ropes, or both, cannot be determined observationally. On the modeling side, the steady-state background solar wind introduces a ${\sim}20^\circ$ offset in the longitude of best agreement and produces unrealistically amplified CME--SIR signatures closer to Earth, preventing a quantitative comparison with STEREO-A. The model also does not resolve the shock and sheath, nor the southward $B_n$ in the upstream solar wind, hence the early phase of the storm is not reproduced. A time-dependent background solar wind \citep[e.g.,][]{baratashvili_2025, samara_2025} may help determine whether the discrepancies between viewpoints arise from the evolution of the CME--SIR interaction or from limitations in the modeled ambient structure.
}

The broader implication is that the geoeffectiveness of a CME cannot always be assessed from its intrinsic properties alone. The structured solar wind through which it propagates can dramatically alter the magnetic field configuration that arrives at Earth. This case demonstrates that an event classified as minor based on its individual characteristics can become severe through heliospheric interactions that occurred beyond the field of view of the instruments used in this study. Although STEREO/HI observations provide coverage at larger heliocentric distances, the limited spatial resolution of these images prevented the unambiguous identification of the CME--SIR interaction for this event. This underscores the need for continuous observational coverage at larger heliocentric distances with improved resolution, and for continuously updating forecasts as new observations become available. For events where such coverage is insufficient, physics-based models can fill the gap by revealing interactions that would otherwise go undetected.

This study also highlights the value of detailed case analyses when forecasts fail. By combining remote-sensing observations, in situ measurements, and MHD simulations from the Sun to 1$\,$au, we identify a heliospheric interaction {that can explain its} unexpected severity: a CME--SIR compression that built up the southward $B_n$ component during propagation. Understanding what was missed in the original forecast, and why, is as important as the forecast itself: it reveals the limitations of current frameworks and points toward the improvements needed, both in observational infrastructure and in the quantification of forecast uncertainty.
 
\begin{acknowledgments}

C.M.\ is supported by the NASA Heliophysics Division, Solar Orbiter Collaboration Office under DPR NNG09EK11I. A.S.\ is supported by an appointment to the NASA Postdoctoral Program at the NASA Goddard Space Flight Center, administered by Oak Ridge Associated Universities under contract with NASA. E.P.\ acknowledges support from NSF's SHINE (Grant AGS-2301403) program as well as NASA's LWS (Grant 80NSSC24K1108), LWS-SC (Grant 80NSSC22K0893), and HSR (Grant 80NSSC25M7101) programs. E.P.'s work was also supported by NASA under Grant 80NSSC26K0467, administered by the NASA Goddard Space Flight Center as part of the Solar Orbiter project. E.S.'s research was supported in part by NASA Solar Orbiter project and NASA Heliophysics Division Space Weather. P.H.\ and R.C.\ are supported by the Office of Naval Research. T.N.-C.\ thanks the support of the Solar Orbiter mission. Simulation results have been provided by the Community Coordinated Modeling Center (CCMC) at Goddard Space Flight Center through their publicly available simulation services (https://ccmc.gsfc.nasa.gov). Solar Orbiter is a mission of international cooperation between the European Space Agency (ESA) and the National Aeronautics and Space Administration (NASA), operated by ESA. The Solar Orbiter Heliospheric Imager (SoloHI) instrument was designed, built, and is now operated by the U.S.\ Naval Research Laboratory with the support of the NASA Heliophysics Division, Solar Orbiter Collaboration Office under DPR NNG09EK11I. 

\end{acknowledgments}

\begin{contribution}

C.M.\ conceptualized the research, conducted the analysis, processed the images, and wrote the manuscript. A.S.\ played a key role in executing the CORHEL-CME simulations, analyzing the model outputs, and comparing them with observations. She also contributed to the manuscript writing. E.P.\ played a key role in generating the synthetic WL images, interpreting the in situ results, and refining the manuscript. E.S.\ provided expertise on the background solar wind and the reproduction of the SIR. P.H., R.C., and T.N-C.\ contributed to the preparation of the manuscript. All authors reviewed and approved the final manuscript.


\end{contribution}

%
\facilities{SDO (AIA), STEREO-A (EUVI, COR2), SOHO (LASCO), SolO (SoloHI), WIND}




\bibliography{sample7}{}

@ARTICLE{hajra_2024,
       author = {{Hajra}, Rajkumar and {Tsurutani}, Bruce Tsatnam and {Lu}, Quanming and {Horne}, Richard B. and {Lakhina}, Gurbax Singh and {Yang}, Xu and {Henri}, Pierre and {Du}, Aimin and {Gao}, Xingliang and {Wang}, Rongsheng and {Lu}, San},
        title = "{The April 2023 SYM-H = {\ensuremath{-}}233 nT Geomagnetic Storm: A Classical Event}",
      journal = {\jgra},
         year = 2024,
        month = oct,
       volume = {129},
       number = {10},
          eid = {e2024JA032986},
        pages = {e2024JA032986},
          doi = {10.1029/2024JA032986},
archivePrefix = {arXiv},
       eprint = {2409.08118},
 primaryClass = {physics.space-ph},
       adsurl = {https://ui.adsabs.harvard.edu/abs/2024JGRA..12932986H}
}

@ARTICLE{paouris_2025,
       author = {{Paouris}, Evangelos and {Vourlidas}, Angelos and {Georgoulis}, Manolis K. and {Hess}, Phillip and {Stenborg}, Guillermo},
        title = "{How the CME on 2023 April 21 Triggered the First Severe Geomagnetic Storm of Solar Cycle 25}",
      journal = {\apj},
         year = 2025,
        month = apr,
       volume = {982},
       number = {2},
          eid = {194},
        pages = {194},
          doi = {10.3847/1538-4357/adb8d3},
archivePrefix = {arXiv},
       eprint = {2503.00705},
 primaryClass = {astro-ph.SR},
       adsurl = {https://ui.adsabs.harvard.edu/abs/2025ApJ...982..194P}
}

@ARTICLE{temmer_2012,
       author = {{Temmer}, Manuela and {Vr{\v{s}}nak}, Bojan and {Rollett}, Tanja and {Bein}, Bianca and {de Koning}, Curt A. and {Liu}, Ying and {Bosman}, Eckhard and {Davies}, Jackie A. and {M{\"o}stl}, Christian and {{\v{Z}}ic}, Tomislav and {Veronig}, Astrid M. and {Bothmer}, Volker and {Harrison}, Richard and {Nitta}, Nariaki and {Bisi}, Mario and {Flor}, Olga and {Eastwood}, Jonathan and {Odstrcil}, Dusan and {Forsyth}, Robert},
        title = "{Characteristics of Kinematics of a Coronal Mass Ejection during the 2010 August 1 CME-CME Interaction Event}",
      journal = {\apj},
         year = 2012,
        month = apr,
       volume = {749},
       number = {1},
          eid = {57},
        pages = {57},
          doi = {10.1088/0004-637X/749/1/57},
archivePrefix = {arXiv},
       eprint = {1202.0629},
 primaryClass = {astro-ph.SR},
       adsurl = {https://ui.adsabs.harvard.edu/abs/2012ApJ...749...57T}
}

@ARTICLE{vemareddy_2024,
       author = {{Vemareddy}, P.},
        title = "{Filament Eruption from Active Region 13283 Leading to a Fast Halo-CME and an Intense Geomagnetic Storm on 2023 April 23}",
      journal = {\apj},
         year = 2024,
        month = feb,
       volume = {961},
       number = {2},
          eid = {199},
        pages = {199},
          doi = {10.3847/1538-4357/ad1662},
archivePrefix = {arXiv},
       eprint = {2312.08790},
 primaryClass = {astro-ph.SR},
       adsurl = {https://ui.adsabs.harvard.edu/abs/2024ApJ...961..199V}
}

@ARTICLE{dungey_1961,
       author = {{Dungey}, J.~W.},
        title = "{Interplanetary Magnetic Field and the Auroral Zones}",
      journal = {\prl},
         year = 1961,
        month = jan,
       volume = {6},
       number = {2},
        pages = {47-48},
          doi = {10.1103/PhysRevLett.6.47},
       adsurl = {https://ui.adsabs.harvard.edu/abs/1961PhRvL...6...47D}
}

@ARTICLE{kamide_1998,
       author = {{Kamide}, Y. and {Baumjohann}, W. and {Daglis}, I.~A. and {Gonzalez}, W.~D. and {Grande}, M. and {Joselyn}, J.~A. and {McPherron}, R.~L. and {Phillips}, J.~L. and {Reeves}, E.~G.~D. and {Rostoker}, G. and {Sharma}, A.~S. and {Singer}, H.~J. and {Tsurutani}, B.~T. and {Vasyliunas}, V.~M.},
        title = "{Current understanding of magnetic storms: Storm-substorm relationships}",
      journal = {\jgr},
         year = 1998,
        month = aug,
       volume = {103},
       number = {A8},
        pages = {17705-17728},
          doi = {10.1029/98JA01426},
       adsurl = {https://ui.adsabs.harvard.edu/abs/1998JGR...10317705K}
}

@ARTICLE{fox_2016,
       author = {{Fox}, N.~J. and {Velli}, M.~C. and {Bale}, S.~D. and {Decker}, R. and {Driesman}, A. and {Howard}, R.~A. and {Kasper}, J.~C. and {Kinnison}, J. and {Kusterer}, M. and {Lario}, D. and {Lockwood}, M.~K. and {McComas}, D.~J. and {Raouafi}, N.~E. and {Szabo}, A.},
        title = "{The Solar Probe Plus Mission: Humanity's First Visit to Our Star}",
      journal = {\ssr},
         year = 2016,
        month = dec,
       volume = {204},
       number = {1-4},
        pages = {7-48},
          doi = {10.1007/s11214-015-0211-6},
       adsurl = {https://ui.adsabs.harvard.edu/abs/2016SSRv..204....7F}
}

@ARTICLE{muller_2020,
       author = {{M{\"u}ller}, D. and {St. Cyr}, O.~C. and {Zouganelis}, I. and {Gilbert}, H.~R. and {Marsden}, R. and {Nieves-Chinchilla}, T. and {Antonucci}, E. and {Auch{\`e}re}, F. and {Berghmans}, D. and {Horbury}, T.~S. and {Howard}, R.~A. and {Krucker}, S. and {Maksimovic}, M. and {Owen}, C.~J. and {Rochus}, P. and {Rodriguez-Pacheco}, J. and {Romoli}, M. and {Solanki}, S.~K. and {Bruno}, R. and {Carlsson}, M. and {Fludra}, A. and {Harra}, L. and {Hassler}, D.~M. and {Livi}, S. and {Louarn}, P. and {Peter}, H. and {Sch{\"u}hle}, U. and {Teriaca}, L. and {del Toro Iniesta}, J.~C. and {Wimmer-Schweingruber}, R.~F. and {Marsch}, E. and {Velli}, M. and {De Groof}, A. and {Walsh}, A. and {Williams}, D.},
        title = "{The Solar Orbiter mission. Science overview}",
      journal = {\aap},
         year = 2020,
        month = oct,
       volume = {642},
          eid = {A1},
        pages = {A1},
          doi = {10.1051/0004-6361/202038467},
archivePrefix = {arXiv},
       eprint = {2009.00861},
 primaryClass = {astro-ph.SR},
       adsurl = {https://ui.adsabs.harvard.edu/abs/2020A&A...642A...1M}
}

@ARTICLE{domingo_1995,
       author = {{Domingo}, V. and {Fleck}, B. and {Poland}, A.~I.},
        title = "{The SOHO Mission: an Overview}",
      journal = {\solphys},
         year = 1995,
        month = dec,
       volume = {162},
       number = {1-2},
        pages = {1-37},
          doi = {10.1007/BF00733425},
       adsurl = {https://ui.adsabs.harvard.edu/abs/1995SoPh..162....1D}
}

@ARTICLE{kaiser_2008,
       author = {{Kaiser}, M.~L. and {Kucera}, T.~A. and {Davila}, J.~M. and {St. Cyr}, O.~C. and {Guhathakurta}, M. and {Christian}, E.},
        title = "{The STEREO Mission: An Introduction}",
      journal = {\ssr},
         year = 2008,
        month = apr,
       volume = {136},
       number = {1-4},
        pages = {5-16},
          doi = {10.1007/s11214-007-9277-0},
       adsurl = {https://ui.adsabs.harvard.edu/abs/2008SSRv..136....5K}
}

@ARTICLE{moestl_2015,
       author = {{M{\"o}stl}, Christian and {Rollett}, Tanja and {Frahm}, Rudy A. and {Liu}, Ying D. and {Long}, David M. and {Colaninno}, Robin C. and {Reiss}, Martin A. and {Temmer}, Manuela and {Farrugia}, Charles J. and {Posner}, Arik and {Dumbovi{\'c}}, Mateja and {Janvier}, Miho and {D{\'e}moulin}, Pascal and {Boakes}, Peter and {Devos}, Andy and {Kraaikamp}, Emil and {Mays}, Mona L. and {Vr{\v{s}}nak}, Bojan},
        title = "{Strong coronal channelling and interplanetary evolution of a solar storm up to Earth and Mars}",
      journal = {\natco},
         year = 2015,
        month = may,
       volume = {6},
          eid = {7135},
        pages = {7135},
          doi = {10.1038/ncomms8135},
archivePrefix = {arXiv},
       eprint = {1506.02842},
 primaryClass = {astro-ph.SR},
       adsurl = {https://ui.adsabs.harvard.edu/abs/2015NatCo...6.7135M}
}

@ARTICLE{linker_2024,
       author = {{Linker}, Jon A. and {Torok}, Tibor and {Downs}, Cooper and {Caplan}, Ronald M. and {Titov}, Viacheslav and {Reyes}, Andres and {Lionello}, Roberto and {Riley}, Pete},
        title = "{CORHEL-CME: An Interactive Tool For Modeling Solar Eruptions}",
         year = 2024,
       journal = {\jphcs},
       volume = {2742},
        month = apr,
          eid = {012012},
        pages = {012012},
          doi = {10.1088/1742-6596/2742/1/012012},
archivePrefix = {arXiv},
       eprint = {2311.03596},
 primaryClass = {astro-ph.SR},
       adsurl = {https://ui.adsabs.harvard.edu/abs/2024JPhCS2742a2012L}
}

@ARTICLE{titov_2018,
       author = {{Titov}, Viacheslav S. and {Downs}, Cooper and {Miki{\'c}}, Zoran and {T{\"o}r{\"o}k}, Tibor and {Linker}, Jon A. and {Caplan}, Ronald M.},
        title = "{Regularized Biot-Savart Laws for Modeling Magnetic Flux Ropes}",
      journal = {\apjl},
         year = 2018,
        month = jan,
       volume = {852},
       number = {2},
          eid = {L21},
        pages = {L21},
          doi = {10.3847/2041-8213/aaa3da},
archivePrefix = {arXiv},
       eprint = {1712.06708},
 primaryClass = {astro-ph.SR},
       adsurl = {https://ui.adsabs.harvard.edu/abs/2018ApJ...852L..21T}
}

@ARTICLE{torok_2018,
       author = {{T{\"o}r{\"o}k}, Tibor and {Downs}, Cooper and {Linker}, Jon A. and {Lionello}, R. and {Titov}, Viacheslav S. and {Miki{\'c}}, Zoran and {Riley}, Pete and {Caplan}, Ronald M. and {Wijaya}, Janvier},
        title = "{Sun-to-Earth MHD Simulation of the 2000 July 14 {\textquotedblleft}Bastille Day{\textquotedblright} Eruption}",
      journal = {\apj},
         year = 2018,
        month = mar,
       volume = {856},
       number = {1},
          eid = {75},
        pages = {75},
          doi = {10.3847/1538-4357/aab36d},
archivePrefix = {arXiv},
       eprint = {1801.05903},
 primaryClass = {astro-ph.SR},
       adsurl = {https://ui.adsabs.harvard.edu/abs/2018ApJ...856...75T}
}

@ARTICLE{hess_2023,
       author = {{Hess}, P. and {Colaninno}, R.~C. and {Vourlidas}, A. and {Howard}, R.~A. and {Stenborg}, G.},
        title = "{SoloHI observations of coronal mass ejections observed by multiple spacecraft}",
      journal = {\aap},
         year = 2023,
        month = nov,
       volume = {679},
          eid = {A149},
        pages = {A149},
          doi = {10.1051/0004-6361/202346907},
       adsurl = {https://ui.adsabs.harvard.edu/abs/2023A&A...679A.149H}
}

@ARTICLE{brueckner_1995,
       author = {{Brueckner}, G.~E. and {Howard}, R.~A. and {Koomen}, M.~J. and {Korendyke}, C.~M. and {Michels}, D.~J. and {Moses}, J.~D. and {Socker}, D.~G. and {Dere}, K.~P. and {Lamy}, P.~L. and {Llebaria}, A. and {Bout}, M.~V. and {Schwenn}, R. and {Simnett}, G.~M. and {Bedford}, D.~K. and {Eyles}, C.~J.},
        title = "{The Large Angle Spectroscopic Coronagraph (LASCO)}",
      journal = {\solphys},
         year = 1995,
        month = dec,
       volume = {162},
       number = {1-2},
        pages = {357-402},
          doi = {10.1007/BF00733434},
       adsurl = {https://ui.adsabs.harvard.edu/abs/1995SoPh..162..357B}
}

@ARTICLE{vourlidas_2014,
       author = {{Vourlidas}, Angelos},
        title = "{The flux rope nature of coronal mass ejections}",
      journal = {\ppcf},
         year = 2014,
        month = jun,
       volume = {56},
       number = {6},
          eid = {064001},
        pages = {064001},
          doi = {10.1088/0741-3335/56/6/064001},
       adsurl = {https://ui.adsabs.harvard.edu/abs/2014PPCF...56f4001V}
}

@ARTICLE{howard_2008,
   author = {{Howard}, R.~A. and {Moses}, J.~D. and {Vourlidas}, A. and {Newmark}, J.~S. and 
	{Socker}, D.~G. and {Plunkett}, S.~P. and {Korendyke}, C.~M. and 
	{Cook}, J.~W. and {Hurley}, A. and {Davila}, J.~M. and {et al.}
	},
    title = "{Sun Earth Connection Coronal and Heliospheric Investigation (SECCHI)}",
  journal = {\ssr},
     year = 2008,
    month = apr,
   volume = 136,
    pages = {67-115},
      doi = {10.1007/s11214-008-9341-4},
}

@ARTICLE{lemen_2012,
   author = {{Lemen}, J.~R. and {Title}, A.~M. and {Akin}, D.~J. and {Boerner}, P.~F. and 
	{Chou}, C. and {Drake}, J.~F. and {Duncan}, D.~W. and {Edwards}, C.~G. and 
	{Friedlaender}, F.~M. and {Heyman}, G.~F. and {et al.}
	},
    title = "{The Atmospheric Imaging Assembly (AIA) on the Solar Dynamics Observatory (SDO)}",
  journal = {\solphys},
     year = 2012,
    month = jan,
   volume = 275,
    pages = {17-40},
      doi = {10.1007/s11207-011-9776-8},
}

@ARTICLE{pesnell_2012,
       author = {{Pesnell}, W. Dean and {Thompson}, B.~J. and {Chamberlin}, P.~C.},
        title = "{The Solar Dynamics Observatory (SDO)}",
      journal = {\solphys},
         year = 2012,
        month = jan,
       volume = {275},
       number = {1-2},
        pages = {3-15},
          doi = {10.1007/s11207-011-9841-3},
       adsurl = {https://ui.adsabs.harvard.edu/abs/2012SoPh..275....3P}
}

@ARTICLE{sahade_2024,
       author = {{Sahade}, Abril and {Vourlidas}, Angelos and {Mac Cormack}, Cecilia},
        title = "{Analysis of Solar Eruptions Deflecting in the Low Corona: Influence of the Magnetic Environment}",
      journal = {\apj},
         year = 2025,
        month = jan,
       volume = {978},
       number = {1},
          eid = {41},
        pages = {41},
          doi = {10.3847/1538-4357/ad96ba},
archivePrefix = {arXiv},
       eprint = {2411.11599},
 primaryClass = {astro-ph.SR},
       adsurl = {https://ui.adsabs.harvard.edu/abs/2025ApJ...978...41S}
}

@ARTICLE{scherrer_2012,
       author = {{Scherrer}, P.~H. and {Schou}, J. and {Bush}, R.~I. and {Kosovichev}, A.~G. and {Bogart}, R.~S. and {Hoeksema}, J.~T. and {Liu}, Y. and {Duvall}, T.~L. and {Zhao}, J. and {Title}, A.~M. and {Schrijver}, C.~J. and {Tarbell}, T.~D. and {Tomczyk}, S.},
        title = "{The Helioseismic and Magnetic Imager (HMI) Investigation for the Solar Dynamics Observatory (SDO)}",
      journal = {\solphys},
         year = 2012,
        month = jan,
       volume = {275},
       number = {1-2},
        pages = {207-227},
          doi = {10.1007/s11207-011-9834-2},
       adsurl = {https://ui.adsabs.harvard.edu/abs/2012SoPh..275..207S}
}

@ARTICLE{patsourakos_2020,
       author = {{Patsourakos}, S. and {Vourlidas}, A. and {T{\"o}r{\"o}k}, T. and {Kliem}, B. and {Antiochos}, S.~K. and {Archontis}, V. and {Aulanier}, G. and {Cheng}, X. and {Chintzoglou}, G. and {Georgoulis}, M.~K. and {Green}, L.~M. and {Leake}, J.~E. and {Moore}, R. and {Nindos}, A. and {Syntelis}, P. and {Yardley}, S.~L. and {Yurchyshyn}, V. and {Zhang}, J.},
        title = "{Decoding the Pre-Eruptive Magnetic Field Configurations of Coronal Mass Ejections}",
      journal = {\ssr},
         year = 2020,
        month = nov,
       volume = {216},
       number = {8},
          eid = {131},
        pages = {131},
          doi = {10.1007/s11214-020-00757-9},
archivePrefix = {arXiv},
       eprint = {2010.10186},
 primaryClass = {astro-ph.SR},
       adsurl = {https://ui.adsabs.harvard.edu/abs/2020SSRv..216..131P}
}

@ARTICLE{howard_2020,
       author = {{Howard}, R.~A. and {Vourlidas}, A. and {Colaninno}, R.~C. and {Korendyke}, C.~M. and {Plunkett}, S.~P. and {Carter}, M.~T. and {Wang}, D. and {Rich}, N. and {Lynch}, S. and {Thurn}, A. and {Socker}, D.~G. and {Thernisien}, A.~F. and {Chua}, D. and {Linton}, M.~G. and {Koss}, S. and {Tun-Beltran}, S. and {Dennison}, H. and {Stenborg}, G. and {McMullin}, D.~R. and {Hunt}, T. and {Baugh}, R. and {Clifford}, G. and {Keller}, D. and {Janesick}, J.~R. and {Tower}, J. and {Grygon}, M. and {Farkas}, R. and {Hagood}, R. and {Eisenhauer}, K. and {Uhl}, A. and {Yerushalmi}, S. and {Smith}, L. and {Liewer}, P.~C. and {Velli}, M.~C. and {Linker}, J. and {Bothmer}, V. and {Rochus}, P. and {Halain}, J. -P. and {Lamy}, P.~L. and {Auch{\`e}re}, F. and {Harrison}, R.~A. and {Rouillard}, A. and {Patsourakos}, S. and {St. Cyr}, O.~C. and {Gilbert}, H. and {Maldonado}, H. and {Mariano}, C. and {Cerullo}, J.},
        title = "{The Solar Orbiter Heliospheric Imager (SoloHI)}",
      journal = {\aap},
         year = 2020,
        month = oct,
       volume = {642},
          eid = {A13},
        pages = {A13},
          doi = {10.1051/0004-6361/201935202},
       adsurl = {https://ui.adsabs.harvard.edu/abs/2020A&A...642A..13H}
}

@ARTICLE{maccormack_2025,
       author = {{Mac Cormack}, Cecilia and {Shaik}, Shaheda B. and {Hess}, Phillip and {Colaninno}, Robin and {Nieves-Chinchilla}, Teresa},
        title = "{A Multi-Viewpoint CME Catalog Based on SoloHI Observed Events}",
      journal = {\solphys},
         year = 2025,
        month = may,
       volume = {300},
       number = {5},
          eid = {54},
        pages = {54},
          doi = {10.1007/s11207-025-02463-7},
archivePrefix = {arXiv},
       eprint = {2503.19060},
 primaryClass = {astro-ph.SR},
       adsurl = {https://ui.adsabs.harvard.edu/abs/2025SoPh..300...54M}
}

@ARTICLE{maccormack_2025b,
       author = {{Mac Cormack}, Cecilia and {Sahade}, Abril and {Vourlidas}, Angelos and {Hess}, Phillip and {Colaninno}, Robin and {Nieves-Chinchilla}, Teresa},
        title = "{Understanding the Complex Morphology of a Coronal Mass Ejection: Multiview Analysis and Numerical Modeling}",
      journal = {\apj},
         year = 2025,
        month = aug,
       volume = {988},
       number = {2},
          eid = {180},
        pages = {180},
          doi = {10.3847/1538-4357/ade311},
archivePrefix = {arXiv},
       eprint = {2506.08820},
 primaryClass = {astro-ph.SR},
       adsurl = {https://ui.adsabs.harvard.edu/abs/2025ApJ...988..180M}
}

@ARTICLE{eastwood_2017,
       author = {{Eastwood}, J.~P. and {Biffis}, E. and {Hapgood}, M.~A. and {Green}, L. and {Bisi}, M.~M. and {Bentley}, R.~D. and {Wicks}, R. and {McKinnell}, L.~A. and {Gibbs}, M. and {Burnett}, C.},
        title = "{The Economic Impact of Space Weather: Where Do We Stand?}",
      journal = {\riska},
         year = 2017,
        month = feb,
       volume = {37},
       number = {2},
        pages = {206-218},
          doi = {10.1111/risa.12765},
       adsurl = {https://ui.adsabs.harvard.edu/abs/2017RiskA..37..206E}
}

@ARTICLE{kumar_2026,
       author = {{Kumar}, Sandeep and {Srivastava}, Nandita and {Banerjee}, Parthib and {Gopalswamy}, Nat},
        title = "{Propagation Characteristics of the April 21, 2023 CME}",
      journal = {JApA},
         year = 2026,
        month = aug,
       volume = {47},
       number = {2},
          eid = {43},
        pages = {43},
          doi = {10.1007/s12036-026-10149-0},
archivePrefix = {arXiv},
       eprint = {2602.16239},
 primaryClass = {astro-ph.SR},
       adsurl = {https://ui.adsabs.harvard.edu/abs/2026arXiv260216239K}
}

@ARTICLE{sarkar_2026,
       author = {{Sarkar}, Ranadeep and {Balan}, Shanmugha and {Srivastava}, Nandita},
        title = "{Sun-to-Earth evolution and space weather impact of the 21 April 2023 CME: validation of a space weather modeling framework}",
      journal = {JApA},
         year = 2026,
        month = jun,
       volume = {47},
       number = {2},
          eid = {33},
        pages = {33},
          doi = {10.1007/s12036-026-10138-3},
       adsurl = {https://ui.adsabs.harvard.edu/abs/2026JApA...47...33S}
}

@article{lugaz_2026,
author = {Lugaz, Noé and Zhuang, Bin and Banu, Sahanaj Aktar and Al-Haddad, Nada and Argall, Matthew R. and Farrugia, Charles J.},
title = {Insights Into the Shape and Sheath Structure of the 23 April 2023 Coronal Mass Ejection From Simultaneous Multi-Spacecraft Measurements},
journal = {\jgra},
volume = {131},
number = {4},
pages = {e2025JA034831},
doi = {https://doi.org/10.1029/2025JA034831},
year = {2026}
}

@ARTICLE{manchester_2017,
       author = {{Manchester}, Ward and {Kilpua}, Emilia K.~J. and {Liu}, Ying D. and {Lugaz}, No{\'e} and {Riley}, Pete and {T{\"o}r{\"o}k}, Tibor and {Vr{\v{s}}nak}, Bojan},
        title = "{The Physical Processes of CME/ICME Evolution}",
      journal = {\ssr},
         year = 2017,
        month = nov,
       volume = {212},
       number = {3-4},
        pages = {1159-1219},
          doi = {10.1007/s11214-017-0394-0},
       adsurl = {https://ui.adsabs.harvard.edu/abs/2017SSRv..212.1159M}
}

@ARTICLE{luhmann_2020,
       author = {{Luhmann}, J.~G. and {Gopalswamy}, N. and {Jian}, L.~K. and {Lugaz}, N.},
        title = "{ICME Evolution in the Inner Heliosphere}",
      journal = {\solphys},
         year = 2020,
        month = apr,
       volume = {295},
       number = {4},
          eid = {61},
        pages = {61},
          doi = {10.1007/s11207-020-01624-0},
       adsurl = {https://ui.adsabs.harvard.edu/abs/2020SoPh..295...61L}
}

@ARTICLE{palmerio_2022,
       author = {{Palmerio}, Erika and {Lee}, Christina O. and {Richardson}, Ian G. and {Nieves-Chinchilla}, Teresa and {Dos Santos}, Luiz F.~G. and {Gruesbeck}, Jacob R. and {Nitta}, Nariaki V. and {Mays}, M. Leila and {Halekas}, Jasper S. and {Zeitlin}, Cary and {Xu}, Shaosui and {Holmstr{\"o}m}, Mats and {Futaana}, Yoshifumi and {Mulligan}, Tamitha and {Lynch}, Benjamin J. and {Luhmann}, Janet G.},
        title = "{CME Evolution in the Structured Heliosphere and Effects at Earth and Mars During Solar Minimum}",
      journal = {\spwea},
         year = 2022,
        month = sep,
       volume = {20},
       number = {9},
          eid = {e2022SW003215},
        pages = {e2022SW003215},
          doi = {10.1029/2022SW003215},
archivePrefix = {arXiv},
       eprint = {2209.05760},
 primaryClass = {astro-ph.SR},
       adsurl = {https://ui.adsabs.harvard.edu/abs/2022SpWea..2003215P}
}

@ARTICLE{harvey_1996,
       author = {{Harvey}, J.~W. and {Hill}, F. and {Hubbard}, R.~P. and {Kennedy}, J.~R. and {Leibacher}, J.~W. and {Pintar}, J.~A. and {Gilman}, P.~A. and {Noyes}, R.~W. and {Title}, A.~M. and {Toomre}, J. and {Ulrich}, R.~K. and {Bhatnagar}, A. and {Kennewell}, J.~A. and {Marquette}, W. and {Patron}, J. and {Saa}, O. and {Yasukawa}, E.},
        title = "{The Global Oscillation Network Group (GONG) Project}",
      journal = {\sci},
         year = 1996,
        month = may,
       volume = {272},
       number = {5266},
        pages = {1284-1286},
          doi = {10.1126/science.272.5266.1284},
       adsurl = {https://ui.adsabs.harvard.edu/abs/1996Sci...272.1284H}
}

@ARTICLE{ogilvie_1997,
       author = {{Ogilvie}, K.~W. and {Desch}, M.~D.},
        title = "{The wind spacecraft and its early scientific results}",
      journal = {\adv},
         year = 1997,
        month = jan,
       volume = {20},
       number = {4-5},
        pages = {559-568},
          doi = {10.1016/S0273-1177(97)00439-0},
       adsurl = {https://ui.adsabs.harvard.edu/abs/1997AdSpR..20..559O}
}

@ARTICLE{barnes_2026,
       author = {{Barnes}, David and {Palmerio}, Erika and {Amerstorfer}, Tanja and {Asvestari}, Eleanna and {Barnard}, Luke and {Bauer}, Maike and {{\v{C}}alogovi{\'c}}, Ja{\v{s}}a and {Cappello}, Greta M. and {Hess}, Phillip and {Kay}, Christina},
        title = "{Polarimetric tomography applied to synthetic multi-spacecraft white-light images: observing coronal mass ejections in 3D}",
      journal = {\mnras},
         year = 2026,
        month = jun,
       volume = {548},
       number = {4},
          eid = {stag818},
        pages = {stag818},
          doi = {10.1093/mnras/stag818},
archivePrefix = {arXiv},
       eprint = {2603.07171},
       adsurl = {https://ui.adsabs.harvard.edu/abs/2026MNRAS.548ag818B}
}

@ARTICLE{baratashvili_2025,
       author = {{Baratashvili}, T. and {Davies}, E. and {Weiler}, E. and {Schmieder}, B. and {D{\'e}moulin}, P. and {Poedts}, S.},
        title = "{Investigating the impact of the dynamic solar wind on the propagation of a coronal mass ejection with two models and multi-spacecraft measurements}",
      journal = {\aap},
         year = 2025,
        month = dec,
       volume = {704},
          eid = {A231},
        pages = {A231},
          doi = {10.1051/0004-6361/202556653},
archivePrefix = {arXiv},
       eprint = {2511.06521},
 primaryClass = {astro-ph.SR},
       adsurl = {https://ui.adsabs.harvard.edu/abs/2025A&A...704A.231B}
}

@ARTICLE{mikic_2018,
       author = {{Miki{\'c}}, Zoran and {Downs}, Cooper and {Linker}, Jon A. and {Caplan}, Ronald M. and {Mackay}, Duncan H. and {Upton}, Lisa A. and {Riley}, Pete and {Lionello}, Roberto and {T{\"o}r{\"o}k}, Tibor and {Titov}, Viacheslav S. and {Wijaya}, Janvier and {Druckm{\"u}ller}, Miloslav and {Pasachoff}, Jay M. and {Carlos}, Wendy},
        title = "{Predicting the corona for the 21 August 2017 total solar eclipse}",
      journal = {\natas},
         year = 2018,
        month = aug,
       volume = {2},
        pages = {913-921},
          doi = {10.1038/s41550-018-0562-5},
       adsurl = {https://ui.adsabs.harvard.edu/abs/2018NatAs...2..913M}
}

@ARTICLE{downs_2025,
       author = {{Downs}, Cooper and {Linker}, Jon A. and {Caplan}, Ronald M. and {Mason}, Emily I. and {Riley}, Pete and {Davidson}, Ryder and {Reyes}, Andres and {Palmerio}, Erika and {Lionello}, Roberto and {Turtle}, James and {Ben-Nun}, Michal and {Stulajter}, Miko M. and {Titov}, Viacheslav S. and {T{\"o}r{\"o}k}, Tibor and {Upton}, Lisa A. and {Attie}, Raphael and {Jha}, Bibhuti K. and {Arge}, Charles N. and {Henney}, Carl J. and {Valori}, Gherardo and {Strecker}, Hanna and {Calchetti}, Daniele and {Germerott}, Dietmar and {Hirzberger}, Johann and {Su{\'a}rez}, David Orozco and {Rodr{\'\i}guez}, Julian Blanco and {Solanki}, Sami K. and {Cheng}, Xin and {Wu}, Sizhe},
        title = "{A near-real-time data-assimilative model of the solar corona}",
      journal = {\sci},
         year = 2025,
        month = jun,
       volume = {388},
       number = {6753},
        pages = {1306-1310},
          doi = {10.1126/science.adq0872},
       adsurl = {https://ui.adsabs.harvard.edu/abs/2025Sci...388.1306D}
}

@ARTICLE{sahade_2026,
       author = {{Sahade}, Abril and {Mac Cormack}, Cecilia and {Vourlidas}, Angelos and {Nieves-Chinchilla}, Teresa and {Downs}, Cooper and {Sasso}, Clementina and {Karpen}, Judith T.},
        title = "{Understanding the Complex Morphology of a Coronal Mass Ejection. II. How Pre-eruptive Conditions Shape Coronal Mass Ejection Evolution}",
      journal = {\apj},
         year = 2026,
        month = jun,
       volume = {1003},
       number = {2},
          eid = {154},
        pages = {154},
          doi = {10.3847/1538-4357/ae64f9},
archivePrefix = {arXiv},
       eprint = {2604.18188},
 primaryClass = {astro-ph.SR},
       adsurl = {https://ui.adsabs.harvard.edu/abs/2026ApJ..1003..154S}
}

@ARTICLE{hayashi_2015,
       author = {{Hayashi}, K. and {Hoeksema}, J.~T. and {Liu}, Y. and {Bobra}, M.~G. and {Sun}, X.~D. and {Norton}, A.~A.},
        title = "{The Helioseismic and Magnetic Imager (HMI) Vector Magnetic Field Pipeline: Magnetohydrodynamics Simulation Module for the Global Solar Corona}",
      journal = {\solphys},
         year = 2015,
        month = may,
       volume = {290},
       number = {5},
        pages = {1507-1529},
          doi = {10.1007/s11207-015-0686-z},
archivePrefix = {arXiv},
       eprint = {1504.05217},
 primaryClass = {astro-ph.SR},
       adsurl = {https://ui.adsabs.harvard.edu/abs/2015SoPh..290.1507H}
}

@ARTICLE{palmerio_2017,
       author = {{Palmerio}, E. and {Kilpua}, E.~K.~J. and {James}, A.~W. and {Green}, L.~M. and {Pomoell}, J. and {Isavnin}, A. and {Valori}, G.},
        title = "{Determining the Intrinsic CME Flux Rope Type Using Remote-sensing Solar Disk Observations}",
      journal = {\solphys},
         year = 2017,
        month = feb,
       volume = {292},
       number = {2},
          eid = {39},
        pages = {39},
          doi = {10.1007/s11207-017-1063-x},
archivePrefix = {arXiv},
       eprint = {1701.08595},
 primaryClass = {astro-ph.SR},
       adsurl = {https://ui.adsabs.harvard.edu/abs/2017SoPh..292...39P}
}

@ARTICLE{lionello_2026,
       author = {{Lionello}, Roberto and {Downs}, Cooper and {Mason}, Emily I. and {Linker}, Jon A. and {Riley}, Pete and {Owens}, Mathew J.},
        title = "{Magnetic Connectivity in the Time-dependent Corona and Heliosphere}",
      journal = {\apjl},
         year = 2026,
        month = apr,
       volume = {1001},
       number = {1},
          eid = {L23},
        pages = {L23},
          doi = {10.3847/2041-8213/ae5791},
archivePrefix = {arXiv},
       eprint = {2603.22440},
 primaryClass = {astro-ph.SR},
       adsurl = {https://ui.adsabs.harvard.edu/abs/2026ApJ..1001L..23L}
}

@ARTICLE{nitta_2021,
       author = {{Nitta}, Nariaki V. and {Mulligan}, Tamitha and {Kilpua}, Emilia K.~J. and {Lynch}, Benjamin J. and {Mierla}, Marilena and {O'Kane}, Jennifer and {Pagano}, Paolo and {Palmerio}, Erika and {Pomoell}, Jens and {Richardson}, Ian G. and {Rodriguez}, Luciano and {Rouillard}, Alexis P. and {Sinha}, Suvadip and {Srivastava}, Nandita and {Talpeanu}, Dana-Camelia and {Yardley}, Stephanie L. and {Zhukov}, Andrei N.},
        title = "{Understanding the Origins of Problem Geomagnetic Storms Associated with ``Stealth'' Coronal Mass Ejections}",
      journal = {\ssr},
         year = 2021,
        month = dec,
       volume = {217},
       number = {8},
          eid = {82},
        pages = {82},
          doi = {10.1007/s11214-021-00857-0},
archivePrefix = {arXiv},
       eprint = {2110.08408},
 primaryClass = {astro-ph.SR},
       adsurl = {https://ui.adsabs.harvard.edu/abs/2021SSRv..217...82N}
}

@ARTICLE{samara_2025,
       author = {{Samara}, Evangelia and {Provornikova}, Elena and {Arge}, C. Nick and {McCubbin}, Andrew J. and {Merkin}, Viacheslav G.},
        title = "{Why Do Solar Wind Models Get It Wrong? Understanding the Capabilities of Time-dependent Solar Wind Simulations}",
      journal = {\apj},
         year = 2025,
        month = dec,
       volume = {994},
       number = {2},
          eid = {236},
        pages = {236},
          doi = {10.3847/1538-4357/ae0b67},
archivePrefix = {arXiv},
       eprint = {2509.20575},
 primaryClass = {astro-ph.SR},
       adsurl = {https://ui.adsabs.harvard.edu/abs/2025ApJ...994..236S}
}

@ARTICLE{alhaddad_2025,
       author = {{Al-Haddad}, Nada and {Lugaz}, No{\'e}},
        title = "{The Magnetic Field Structure of Coronal Mass Ejections: A More Realistic Representation}",
      journal = {\ssr},
         year = 2025,
        month = feb,
       volume = {221},
       number = {1},
          eid = {12},
        pages = {12},
          doi = {10.1007/s11214-025-01138-w},
       adsurl = {https://ui.adsabs.harvard.edu/abs/2025SSRv..221...12A}
}

@ARTICLE{owens_2017,
       author = {{Owens}, M.~J. and {Lockwood}, M. and {Barnard}, L.~A.},
        title = "{Coronal mass ejections are not coherent magnetohydrodynamic structures}",
      journal = {\natsr},
         year = 2017,
        month = jun,
       volume = {7},
          eid = {4152},
        pages = {4152},
          doi = {10.1038/s41598-017-04546-3},
       adsurl = {https://ui.adsabs.harvard.edu/abs/2017NatSR...7.4152O}
}

@ARTICLE{scolini_2020,
       author = {{Scolini}, Camilla and {Chan{\'e}}, Emmanuel and {Temmer}, Manuela and {Kilpua}, Emilia K.~J. and {Dissauer}, Karin and {Veronig}, Astrid M. and {Palmerio}, Erika and {Pomoell}, Jens and {Dumbovi{\'c}}, Mateja and {Guo}, Jingnan and {Rodriguez}, Luciano and {Poedts}, Stefaan},
        title = "{CME-CME Interactions as Sources of CME Geoeffectiveness: The Formation of the Complex Ejecta and Intense Geomagnetic Storm in 2017 Early September}",
      journal = {\apjs},
         year = 2020,
        month = feb,
       volume = {247},
       number = {1},
          eid = {21},
        pages = {21},
          doi = {10.3847/1538-4365/ab6216},
archivePrefix = {arXiv},
       eprint = {1911.10817},
 primaryClass = {astro-ph.SR},
       adsurl = {https://ui.adsabs.harvard.edu/abs/2020ApJS..247...21S}
}

@INPROCEEDINGS{demoulin_2010,
       author = {{D{\'e}moulin}, Pascal},
        title = "{Interaction of ICMEs with the Solar Wind}",
    booktitle = {Twelfth International Solar Wind Conference},
         year = 2010,
       editor = {{Maksimovic}, M. and {Issautier}, K. and {Meyer-Vernet}, N. and {Moncuquet}, M. and {Pantellini}, F.},
       series = {American Institute of Physics Conference Series},
       volume = {1216},
        month = mar,
    publisher = {AIP},
        pages = {329-334},
          doi = {10.1063/1.3395866},
       adsurl = {https://ui.adsabs.harvard.edu/abs/2010AIPC.1216..329D}
}

@ARTICLE{ruffenach_2012,
       author = {{Ruffenach}, A. and {Lavraud}, B. and {Owens}, M.~J. and {Sauvaud}, J.-A. and {Savani}, N.~P. and {Rouillard}, A.~P. and {D{\'e}moulin}, P. and {Foullon}, C. and {Opitz}, A. and {Fedorov}, A. and {Jacquey}, C.~J. and {G{\'e}not}, V. and {Louarn}, P. and {Luhmann}, J.~G. and {Russell}, C.~T. and {Farrugia}, C.~J. and {Galvin}, A.~B.},
        title = "{Multispacecraft observation of magnetic cloud erosion by magnetic reconnection during propagation}",
      journal = {\jgra},
         year = 2012,
        month = sep,
       volume = {117},
       number = {A9},
          eid = {A09101},
        pages = {A09101},
          doi = {10.1029/2012JA017624},
       adsurl = {https://ui.adsabs.harvard.edu/abs/2012JGRA..117.9101R}
}

@ARTICLE{alhaddad_2019,
       author = {{Al-Haddad}, N. and {Poedts}, S. and {Roussev}, I. and {Farrugia}, C.~J. and {Yu}, W. and {Lugaz}, N.},
        title = "{The Magnetic Morphology of Magnetic Clouds: Multi-spacecraft Investigation of Twisted and Writhed Coronal Mass Ejections}",
      journal = {\apj},
         year = 2019,
        month = jan,
       volume = {870},
       number = {2},
          eid = {100},
        pages = {100},
          doi = {10.3847/1538-4357/aaf38d},
       adsurl = {https://ui.adsabs.harvard.edu/abs/2019ApJ...870..100A}
}

@ARTICLE{isavnin_2013,
       author = {{Isavnin}, A. and {Vourlidas}, A. and {Kilpua}, E.~K.~J.},
        title = "{Three-Dimensional Evolution of Erupted Flux Ropes from the Sun (2 - 20 R $_{☉}$) to 1 AU}",
      journal = {\solphys},
         year = 2013,
        month = may,
       volume = {284},
       number = {1},
        pages = {203-215},
          doi = {10.1007/s11207-012-0214-3},
archivePrefix = {arXiv},
       eprint = {1211.2108},
 primaryClass = {astro-ph.SR},
       adsurl = {https://ui.adsabs.harvard.edu/abs/2013SoPh..284..203I}
}

@ARTICLE{kilpua_2019,
       author = {{Kilpua}, E.~K.~J. and {Lugaz}, N. and {Mays}, M.~L. and {Temmer}, M.},
        title = "{Forecasting the Structure and Orientation of Earthbound Coronal Mass Ejections}",
      journal = {\spwea},
         year = 2019,
        month = apr,
       volume = {17},
       number = {4},
        pages = {498-526},
          doi = {10.1029/2018SW001944},
       adsurl = {https://ui.adsabs.harvard.edu/abs/2019SpWea..17..498K}
}

@ARTICLE{palmerio_2018,
       author = {{Palmerio}, E. and {Kilpua}, E.~K.~J. and {M{\"o}stl}, C. and {Bothmer}, V. and {James}, A.~W. and {Green}, L.~M. and {Isavnin}, A. and {Davies}, J.~A. and {Harrison}, R.~A.},
        title = "{Coronal Magnetic Structure of Earthbound CMEs and In Situ Comparison}",
      journal = {\spwea},
         year = 2018,
        month = may,
       volume = {16},
       number = {5},
        pages = {442-460},
          doi = {10.1002/2017SW001767},
archivePrefix = {arXiv},
       eprint = {1803.04769},
 primaryClass = {astro-ph.SR},
       adsurl = {https://ui.adsabs.harvard.edu/abs/2018SpWea..16..442P}
}

@ARTICLE{gopalswamy_2022,
       author = {{Gopalswamy}, N. and {Yashiro}, S. and {Akiyama}, S. and {Xie}, H. and {M{\"a}kel{\"a}}, P. and {Fok}, M.-C. and {Ferradas}, C.~P.},
        title = "{What Is Unusual About the Third Largest Geomagnetic Storm of Solar Cycle 24?}",
      journal = {\jgra},
         year = 2022,
        month = aug,
       volume = {127},
       number = {8},
          eid = {e30404},
        pages = {e30404},
          doi = {10.1029/2022JA030404},
archivePrefix = {arXiv},
       eprint = {2207.11630},
 primaryClass = {astro-ph.SR},
       adsurl = {https://ui.adsabs.harvard.edu/abs/2022JGRA..12730404G}
}

@ARTICLE{lugaz_2017,
       author = {{Lugaz}, No{\'e} and {Temmer}, Manuela and {Wang}, Yuming and {Farrugia}, Charles J.},
        title = "{The Interaction of Successive Coronal Mass Ejections: A Review}",
      journal = {\solphys},
         year = 2017,
        month = apr,
       volume = {292},
       number = {4},
          eid = {64},
        pages = {64},
          doi = {10.1007/s11207-017-1091-6},
archivePrefix = {arXiv},
       eprint = {1612.02398},
 primaryClass = {physics.space-ph},
       adsurl = {https://ui.adsabs.harvard.edu/abs/2017SoPh..292...64L}
}

@ARTICLE{nieves_2012,
       author = {{Nieves-Chinchilla}, T. and {Colaninno}, R. and {Vourlidas}, A. and {Szabo}, A. and {Lepping}, R.~P. and {Boardsen}, S.~A. and {Anderson}, B.~J. and {Korth}, H.},
        title = "{Remote and in situ observations of an unusual Earth-directed coronal mass ejection from multiple viewpoints}",
      journal = {Journal of Geophysical Research (Space Physics)},
         year = 2012,
        month = jun,
       volume = {117},
       number = {A6},
          eid = {A06106},
        pages = {A06106},
          doi = {10.1029/2011JA017243},
       adsurl = {https://ui.adsabs.harvard.edu/abs/2012JGRA..117.6106N}
}

@ARTICLE{nieves_2018,
       author = {{Nieves-Chinchilla}, T. and {Vourlidas}, A. and {Raymond}, J.~C. and {Linton}, M.~G. and {Al-haddad}, N. and {Savani}, N.~P. and {Szabo}, A. and {Hidalgo}, M.~A.},
        title = "{Understanding the Internal Magnetic Field Configurations of ICMEs Using More than 20 Years of Wind Observations}",
      journal = {\solphys},
         year = 2018,
        month = feb,
       volume = {293},
       number = {2},
          eid = {25},
        pages = {25},
          doi = {10.1007/s11207-018-1247-z},
       adsurl = {https://ui.adsabs.harvard.edu/abs/2018SoPh..293...25N}
}

@ARTICLE{nieves_2013,
       author = {{Nieves-Chinchilla}, T. and {Vourlidas}, A. and {Stenborg}, G. and {Savani}, N.~P. and {Koval}, A. and {Szabo}, A. and {Jian}, L.~K.},
        title = "{Inner Heliospheric Evolution of a ``Stealth'' CME Derived from Multi-view Imaging and Multipoint in Situ observations. I. Propagation to 1 AU}",
      journal = {\apj},
         year = 2013,
        month = dec,
       volume = {779},
       number = {1},
          eid = {55},
        pages = {55},
          doi = {10.1088/0004-637X/779/1/55},
archivePrefix = {arXiv},
       eprint = {1311.6895},
 primaryClass = {astro-ph.SR},
       adsurl = {https://ui.adsabs.harvard.edu/abs/2013ApJ...779...55N}
}

@ARTICLE{gonzalez_1994,
       author = {{Gonzalez}, W.~D. and {Joselyn}, J.~A. and {Kamide}, Y. and {Kroehl}, H.~W. and {Rostoker}, G. and {Tsurutani}, B.~T. and {Vasyliunas}, V.~M.},
        title = "{What is a geomagnetic storm?}",
      journal = {\jgr},
         year = 1994,
        month = apr,
       volume = {99},
       number = {A4},
        pages = {5771-5792},
          doi = {10.1029/93JA02867},
       adsurl = {https://ui.adsabs.harvard.edu/abs/1994JGR....99.5771G}
}

@ARTICLE{chen_1996,
       author = {{Chen}, James},
        title = "{Theory of prominence eruption and propagation: Interplanetary consequences}",
      journal = {\jgr},
         year = 1996,
        month = dec,
       volume = {101},
       number = {A12},
        pages = {27499-27520},
          doi = {10.1029/96JA02644},
       adsurl = {https://ui.adsabs.harvard.edu/abs/1996JGR...10127499C}
}

@ARTICLE{low_1984,
       author = {{Low}, B.~C.},
        title = "{Self-Similar Magnetohydrodynamics - Part Four - the Physics of Coronal Transients}",
      journal = {\apj},
         year = 1984,
        month = jun,
       volume = {281},
        pages = {392},
          doi = {10.1086/162110},
       adsurl = {https://ui.adsabs.harvard.edu/abs/1984ApJ...281..392L}
}

@ARTICLE{dasso_2006,
       author = {{Dasso}, S. and {Mandrini}, C.~H. and {D{\'e}moulin}, P. and {Luoni}, M.~L.},
        title = "{A new model-independent method to compute magnetic helicity in magnetic clouds}",
      journal = {\aap},
         year = 2006,
        month = aug,
       volume = {455},
       number = {1},
        pages = {349-359},
          doi = {10.1051/0004-6361:20064806},
       adsurl = {https://ui.adsabs.harvard.edu/abs/2006A&A...455..349D}
}

@ARTICLE{burton_1975,
       author = {{Burton}, R.~K. and {McPherron}, R.~L. and {Russell}, C.~T.},
        title = "{An empirical relationship between interplanetary conditions and Dst}",
      journal = {\jgr},
         year = 1975,
        month = nov,
       volume = {80},
       number = {31},
        pages = {4204},
          doi = {10.1029/JA080i031p04204},
       adsurl = {https://ui.adsabs.harvard.edu/abs/1975JGR....80.4204B}
}
\bibliographystyle{aasjournalv7}



\end{document}